\documentclass[superscriptaddress,aps,twocolumn,longbibliography]{revtex4-2}
\usepackage{amsmath,amssymb}
\usepackage{graphicx}
\usepackage{dcolumn}
\usepackage[bb=boondox]{mathalfa}
\usepackage{lipsum}
\usepackage{enumitem}
\usepackage{float}
\usepackage{bm,dsfont}
\usepackage{multirow}
\usepackage{simpler-wick}
\usepackage{cuted}
\usepackage{color}
\usepackage{hyperref}
\usepackage{lipsum}
\usepackage{natbib}
\hypersetup{
colorlinks = true,
linkcolor = [rgb]{0.70,0.13,0.13},
citecolor = [rgb]{0.13,0.55,0.13},
urlcolor = [rgb]{0.25, 0.41, 0.88}}

\newcommand{\eqnref}[1]{Eq.\,\eqref{#1}}
\newcommand{\figref}[1]{Fig.\,\ref{#1}}

\usepackage[mathlines]{lineno}

\begin{document}

\title{Differentiable Programming of Isometric Tensor Networks}
\author{Chenhua Geng}
\thanks{xwkgch@issp.u-tokyo.ac.jp}
\affiliation{Institute for Solid State Physics, The University of Tokyo, Kashiwa, Chiba 277-8581, Japan}
\author{Hong-Ye Hu}
\thanks{hyhu@ucsd.edu}
\affiliation{Department of Physics, University of California San Diego, La Jolla, CA 92093, USA}
\author{Yijian Zou}
\thanks{zouy@stanford.edu}
\affiliation{Stanford Institute for Theoretical Physics, Stanford University, Palo Alto, CA 94305, USA}

\date{\today}

\begin{abstract}

Differentiable programming is a new programming paradigm which enables large scale optimization through automatic calculation of gradients  also known as auto-differentiation. This concept emerges from deep learning, and has also been generalized to tensor network optimizations. Here, we extend the differentiable programming to tensor networks with isometic constraints with applications to multiscale entanglement renormalization ansatz (MERA) and tensor network renormalization (TNR). By introducing several gradient-based optimization methods for the isometric tensor network and comparing with Evenbly-Vidal method, we show that auto-differentiation has a better performance for both stability and accuracy. We numerically tested our methods on 1D critical quantum Ising spin chain and 2D classical Ising model. We calculate the ground state energy for the 1D quantum model and internal energy for the classical model, and scaling dimensions of scaling operators and find they all agree with the theory well.


\end{abstract}

\pacs{Valid PACS appear here}

\maketitle


\section{Introduction}

Tensor network has been a powerful tool to study quantum many-body systems and classical statistical-mechanical models both theoretically \cite{random_tensor_network,holographic_tensor_network} and numerically \cite{Verstraete2004cf,PhysRevLett.101.110501,PhysRevLett.91.147902,RevModPhys.77.259,PhysRevX.9.021040,doi:10.1080/14789940801912366}. And recently, it has been proposed as an alternative tool for (quantum) machine learning tasks, both for supervised learning \cite{2016arXiv160503795N,NIPS2016_5314b967,2018arXiv180605964G,2020arXiv200706082M,PhysRevB.101.075135,2019arXiv190606329E,2020arXiv200909932C,2021arXiv210306872L}, and unsupervised learning \cite{2019PhRvB..99o5125C,PhysRevX.8.031012,e21121236,2019arXiv190703741G, gao2021enhancing,PhysRevA.98.062324,2019PhRvB..99o5125C}. In the family of tensor networks, the multi-scale entanglement renormalization ansatz (MERA) and tensor network renormalization (TNR) are important tensor networks which are inspired by the idea of renormalization group (RG). They contain tensors that are to be optimized under isometric constraints, where some tensors are restricted to be isometric.

RG \cite{Kadanoff1966,WILSON197475,wilson1983} plays an important role in modern condensed matter physics and high-energy physics.  What lies in the heart of RG is the coarse graining procedure that changes the scale the system. Under RG, microscopic models flow to different fixed points that distinguish different macroscopic phases of matter. More recently, the concept of RG also finds applications in machine learning and artificial intelligence (AI) \cite{2020arXiv201000029H,PhysRevLett.121.260601,koch2018mutual,2021arXiv210111306L,2019arXiv190506352E}. Traditionally, RG is performed in the Fourier space, which involves various approximations. Furthermore, it has been shown that RG can be performed in real space using tensor networks.

In the context of quantum many-body physics, MERA and its variations \cite{PhysRevLett.101.110501,evenbly2009algorithms,vidal2007entanglement,PhysRevLett.100.240603,evenbly2015tensor} are tensor networks which perform real-space RG on quantum states.
Due to the specific structure, MERA is especially suitable for describing systems with scale invariance such as quantum critical systems.
On the other hand, for a classical statistical mechanical model, TNR \cite{evenbly2015tensor,evenbly2017algorithms} is a tensor network algorithm that performs real-space RG for the tensor network that represents the partition function of this model. 
Compared to  tensor renormalization group (TRG) \cite{levin2007tensor} which also performs real-space RG, TNR resolves the computational breakdown of TRG for critical systems. 
Additionally, it is known that TNR is closely related to MERA as formalized in Ref. \cite{evenbly2015yield}.

From the perspective of theoretical physics, MERA and TNR can be used to extract universal information of critical systems, which are in a close relationship with conformal field theory (CFT) \cite{PhysRevA.79.040301,PhysRevD.95.066004}.
Recently, another variation of MERA using neural network was proposed and had been applied to simulation of quantum field theories, and finding holographic duality based on field theory actions. \cite{PhysRevLett.121.260601,PhysRevResearch.2.023369}.

From the perspective of numerical studies, one of the central problems to tensor network is the optimization of tensors for various systems. Conventional optimization algorithms are designed manually and separately for different problems and different tensor networks structures. Recently, differentiable programming, as a novel programming paradigm, has been proposed for the optimization problems of tensor networks based on gradient optimization, which is extensively used in deep learning and artificial intelligence. Compared with traditional gradient estimation, such as finite difference method, automatic differentiation has the merits such as high accuracy and low computational complexity for calculating gradients for many parameters.
It provides a unified and elegant solution to many optimization problems by combining the well-developed automatic differentiation frameworks like PyTorch \cite{paszke2019pytorch} and TensorFlow \cite{199317tensorflow}. Recently, it has been successfully applied to TRG and the optimization of projected entangled pair states (PEPS) \cite{PhysRevX.9.031041}. With the help of differentiable programming researchers can focus on the part of tensor contraction in the algorithm without worrying about detailed and complicated optimization algorithms.
However, tensor networks which possess isometric constraints, such as MERA and TNR cannot be optimized using differentiable programming directly. 

Recently, Hauru et al. \cite{hauru2020riemannian} initiated the use of Riemannian optimization to tensor networks with isometric tensors.
Specifically, they have applied preconditioned conjugate gradient method to MERA and matrix product states.
In this work, we use a modified gradient search and show that auto-differentiation can be applied to MERA and TNR as well.
We explicitly construct the computation graphs for MERA and TNR algorithms.
Further, we discuss the gradient-based optimization methods and find that the combination of Evenbly-Vidal and gradient-based methods generally have a better performance than either of them. 
Taking 1D quantum and 2D classical Ising model as examples, we obtain the ground state energy for the 1D quantum model and internal energy for the 2D classical model with high accuracy.
We also obtain scaling dimensions of scaling operators at the critical point within the differentiable programming framework.




This paper is organized as follows: In Sec. II we review the idea of differentiable programming, including automatic differentiation, computation graphs and gradient-based optimization.
We introduce our new methods for optimizing isometric tensors within differentiable programming framework.
In Sec. III and IV we briefly review MERA and TNR respectively, and show the results computed using differentiable programming. Finally we make a summary about differentiable programming and give our outlook for future research directions in Sec. V.

\section{Differentiable Programming}

The characteristics of differentiable programming is automatic differentiation which computes the derivative information automatically by the well-developed frameworks. 
We first review the core concepts of differentiable programming and investigate how the derivative information are automatically computed for tensor networks.
Then we discuss how to optimize a tensor network, especially with isometric constraints, by these derivative information.

\subsection{Computation graphs}
\begin{figure}[htbp]
    \centering
    \includegraphics[width=0.4\textwidth]{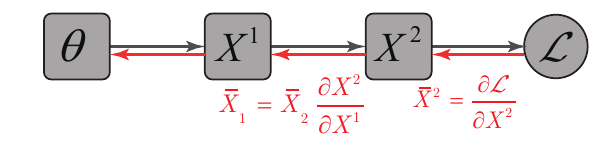}
    \caption{An example of the chain graph. The forward and backward propagation are indicated by black and red arrows.}
    \label{fig.chain}
\end{figure}

Computation graphs are central to the automatic differentiation which present how the derivatives are computed with respect to intermediate variables by the chain rule
\begin{equation}
    \frac{\partial \mathcal{L}}{\partial \theta} = \frac{\partial \mathcal{L}}{\partial X^n} \frac{\partial X^n}{\partial X^{n-1}}...\frac{\partial X^1}{\partial \theta},
\label{chain}
\end{equation}
where $\mathcal{L}$ is some general objective function used for the optimization, $\theta$ is a general trainable variable and $X^i$ are intermediate variables. A computation graph is a directed acyclic graph where a node represents a tensor which stores the data and a link indicates the dependence of data flow during the computation process.
The simplest computation graph is the chain graph characterized by \eqnref{chain}, see \figref{fig.chain}.

In differentiable programming, all the data belongs to tensors which are represented by higher-order arrays. 
The data together with trainable variable $\theta$ flows along the computation graph with a series of intermediate tensors to obtain the final output results.
This computation process is referred to as the forward propagation.

In most application scenarios, there are some tensors to be determined and optimized in the computation graph.
For this purpose, the output results are usually compared with the expected output data to obtain a scalar, known as loss function, which evaluates the quality of the computation process. 
From the loss function, the derivatives with respect to the intermediate tensors are computed along the reversed computation graph, which is the so-called backward propagation.
Once a computation graph has been set up, one can directly obtain the derivative tensors of output with respect to the input or other intermediate tensors through backward propagation.
Using the derivative information of the intermediate tensors, the parameters in the computation graph can be optimized by various optimization methods which will be discussed later. 

For a general computation graph, the computation of the derivative of a node need to sum over all contributions from its child nodes
\begin{equation}
    \bar{X}^i = \sum_{j: \text{child of } i} \bar{X}^j \frac{\partial X^j}{\partial X^i},
\end{equation}
where the adjoint variable $\bar{X} = \partial \mathcal{L} / \partial X$ is defined as the derivative of the final output $\mathcal{L}$ with respect to the variable $X$ as shown in \figref{fig.chain}.
If a node $X$ has several different paths to flow and affect the final output, the derivative $\bar{X}$ will account all the contribution among these paths.

In the conventional tensor network computation, it usually involves computing the environment of a tensor which is also a tensor obtained by computing a fully contracted network without this tensor.
These environment tensors are used to optimize the tensor network, but it is laborious to draw and determine the details of the contraction graph of the environment tensor by hands.
Besides, because of the tensor contraction operation with a large amount of tensors, computing the environment of a tensor is costly for the computer.
We note that the derivation with respect to a tensor is computed by removing the tensor and compute the contraction of the remnant tensor network, which is exactly the derivative of a tensor is exactly the environment of the tensor so long as the computation graph is constructed properly.
The observation above motivates the application of automatic differentiation to optimizing tensor networks in order to get rid of the tedious environments computation. The differentiable programming can be easily implemented using modern machine learning frameworks, like PyTorch, TensorFlow. Furthermore, these frameworks provide easy-to-use interfaces with high performance computing units, such as Graphics Processing Units (GPUs) and Tensor Processing Units (TPUs), which remarkably boost the computation speed.


\subsection{Gradient-based Optimization} \label{sub.gradopt}
Having established the computation graph, we need to determine the optimization method used in differentiable programming.
Given the input data $\mathcal{D}$, the optimization problem is to find an optimal mapping $f(\mathcal{D})$ with parameters $X$ to minimize the loss function 
\begin{equation}
    \min _{X} \mathcal{L} (\mathcal{D}, f_{X}(\mathcal{D})) .
\end{equation}

Gradient descent method is the earliest and most common optimization method.
The idea of gradient descent method is to iteratively update the parameter $X$ by subtracting the gradient of the parameter $\bar{X}$ timed by a factor $\eta$, known as learning rate, 
\begin{equation}
    X_{t+1} \leftarrow X_{t} - \eta \bar{X}_{t} .
\end{equation}


Some studies have found that in many cases the difficulty of optimization comes from the ``saddle point" \cite{pascanu2013difficulty} where the slope is positive in some directions and negative in another directions.
There are several ways to help the optimization escape saddle points.

For example, the momentum \cite{polyak1964some} is introduced in gradient descent to simulate the inertia of optimization process. 
In the momentum method the historical influence is taken into account 
\begin{align}
    M_{t+1} & \leftarrow \beta_m M_t + \eta \bar{X_t} ,\label{momentum1} \\ 
    X_{t+1} & \leftarrow X_t - M_{t+1} ,
\end{align}
where $\beta_m\leq 1$ is the so-called momentum factor. 
With benefiting from the extra variable $M$, the momentum method can help speed up the convergence and get away from saddle points.

Another direction of improving gradient descent is to adjust the learning rate dynamically. 
For instance, RMSprop \cite{tieleman2012rmsprop} uses the historical decayed gradients accumulated up to adjust the learning rate automatically by
\begin{align}
    v_{t+1} & \leftarrow \sqrt{\beta_v v_t + (1-\beta_v) (\bar{X_t})^2} , \\ 
    X_{t+1} & \leftarrow X_t - \frac{\eta}{v_{t+1}} \bar{X_t} ,
\end{align}
where $\beta_v$ is a decay factor.

\subsection{Riemannian Optimization for isometric tensors\label{sec:Rieman}}
In many physics problems, some of the tensors and parameters are required to be satisfied isometric constraints $X^TX=I$ where $X$ is the parameter to be optimized of the form of matrices and tensors.
The isometric constraints for tensor contraction is illustrated in \figref{fig.constraints} and \figref{fig.TNRconstraints}.
In this paper we assume the parameter $X$ is real and $X^T$ denotes the transpose of $X$. The generalization to complex cases is straight forward.

However, the isometric constraints, also referred as  (semi-) orthogonal or (semi-) unitary constraints, obstruct the direct application of the optimization methods in the previous subsection. 
There are two ways to impose the constraints during the optimization.

One is the soft-constraint optimization, which allows the isometries and unitaries away from the constraints and add the deviation from the constraints into the loss function. This is known as the Lagrange multiplier method \cite{bertsekas2014constrained}.
Then it becomes an unconstrained problem which the gradients of tensors are simply derivatives of tensors. 
We can directly use the inbuilt optimizers of the deep learning frameworks to minimize the modified loss function which is composed of the energy and deviation, see Appendix \ref{app.soft}.

The other is the hard-constraint optimizations, which restrict the isometries and unitaries to be isometric. This corresponds to the optimization problems on the Stiefel manifold \cite{tagare2011notes,wen2013feasible}. In this case, the gradients of tensors should be confined in the Stiefel manifold, being differ from the simply derivatives of tensors which is the gradient tensors of the whole Euclid space. In the following, we focus on the hard-constraint optimization.

The hard-constraint optimization requires the tensors to be optimized satisfying the constraints during the whole computation process.
A widely used strategy for solving the problem is the Riemannian optimization.
Various methods based on Riemannian optimization have been studied extensively, including polar decomposition \cite{absil2012projection}, QR decomposition \cite{kaneko2012empirical}, Cayley transform \cite{nishimori2005learning,jiang2015framework,zhu2017riemannian,wen2013feasible, li2020efficient}.
The applications of Riemannian optimization have also been studied in physics like quantum control and quantum technologies \cite{pechen2008control, oza2009optimization}.

Basically, the Riemannian optimization includes two steps:
(1) Find the gradient vector in the tangent space of the current point on Stiefel manifold.
(2) Find the descent direction and ensuring the new points on the manifold.

Specifically, assume the tensor $X$ is one of the tensors to be optimized with the constraints.
Define the set $\{ X \in \mathbb{R} ^{n \times p}: X^T X = I, n\geqslant p \}$ as the Stiefel manifold with the dimension $np-\frac{1}{2}p(p+1)$.
And define the tangent space at the point $X$ as $\mathcal{T}_X = \{ Z \in \mathbb{R} ^{n \times p}: Z^T X + X^T Z = 0 \}$.
Now the problem becomes to found a point on the Stiefel manifold with the minimum loss function $\mathcal{L}$
\begin{equation}
    \min _{X \in \mathbb{R} ^{n \times p}} \mathcal{L} (X) \quad \text{s.t.} \, X^T X = I .
\end{equation}

For the first step, the inner product in the tangent space $\mathcal{T}_X$ should be defined to obtain the gradient $G_X \in \mathcal{T}_X$. 
Let $Z_1, Z_2 \in \mathcal{T}_X$, we can define the Euclidean inner product as $ \left\langle Z_1,Z_2 \right\rangle_e = \text{tr} (Z_1^T Z_2) $.
But it is more widely used to define a more natural choice which is the canonical inner product $ \left\langle  Z_1,Z_2 \right\rangle _c = \text{tr} (Z_1^T (I-\frac{1}{2}X X^T) Z_2)  $.
In the following we choose the canonical inner product. 
It can be proved \cite{wen2013feasible} that the gradient $G_X$ on the manifold is 
\begin{equation}
    G_X = AX = \bar{X} - \frac{1}{2}(X X^T \bar{X} + X \bar{X}^T X),
    \label{gradient}
\end{equation}
where $A = \bar{X} X^T - X \bar{X}^T + \frac{1}{2}X(\bar{X}^T X - X^T \bar{X})X^T$. 
In other words, the gradient $G_X$ is obtained by projecting the derivative $\bar{X}$ onto the tangent space $\mathcal{T}_X$ of Stiefel manifold. 

For the second step, the so-called operation retraction plays an important role.
Generally speaking, there are two classes of retraction to keep the updated point on the manifold: projection-like and geodesic-like schemes.
The projection-like schemes preserve the constraint by projecting a point into the manifold such as QR decomposition, while the geodesic-like schemes preserve the constraint by moving a point along the geodesic or quasi-geodesic line such as the Cayley transform.

\subsubsection{projection-like schemes}
The point $X$ on the manifold moves along the gradient vector a short distance to the new point $X - \eta G_X$, where $\eta$ is a tunable parameter representing to the learning rate.
But generally, the new point $X - \eta G_X$ is not the point on the manifold.
We need to project the point $X - \eta G_X$ onto Stiefel manifold by various methods.

One way is the QR-decomposition-type retraction by noting that the Q factor of QR decomposition is orthogonal.
Then the update equation is 
\begin{align}
    QR &= X_t - \frac{1}{2}\eta G_X , \label{qr}\\
    X_{t+1} &= Q .
\end{align}

Another way is to use SVD as retraction.
If the SVD of a matrix $X \in \mathbb{R} ^{n \times p}$ is $X = U \Sigma V^T$, then the projection map onto Stiefel manifold is $\pi(X) = U I_{n \times p} V^T$ \cite{manton2002optimization}.
Then the update equation is 
\begin{align}
    U \Sigma V^T &= X_t - \eta G_X , \\
    X_{t+1} &= U I_{n \times p} V^T .
    \label{svd}
\end{align}

\subsubsection{geodesic-like schemes}
The point $X$ on the manifold moves along a single parameter curve $Y(\eta)$ such that the curve is on the Stiefel manifold, i.e. $Y(\eta)^T Y(\eta) = I$, and the derivative of the curve at origin is the gradient vector $Y(0)'=-G_X$.

One of the choice of the curve is Cayley transform
\begin{equation}
    Y(\eta) =\left( I + \frac{\eta}{2} A\right) ^{-1} \left(I - \frac{\eta}{2} A\right)  X ,
    \label{Cayley0}
\end{equation}
where $A$ is the matrix in \eqnref{gradient}.
Then update equation is
\begin{equation}
    X_{t+1} = \left(I + \frac{\eta}{2} A_t\right)^{-1} \left(I - \frac{\eta}{2} A_t\right)  X_t .
    \label{Cayley}
\end{equation}

Note that the inverting a $n \times n$ matrix $I + \frac{\eta}{2} A$ is costly.
In Appendix \ref{app.Cayleyreduce}, we show that the computation expense can be reduced by Sherman-Morrison-Woodbury formula or an iteration method.

\textbf{Algorithm improvement}: 
In order to improve the optimization performance, We apply the optimization techniques introduced in Sect. \ref{sub.gradopt} to our Riemannian optimization methods.
Generally, both the accuracy and optimization speed could be improved by introducing the dynamic momentum and adaptive learning rate techniques \cite{li2020efficient}.

To be specific, instead of directly using the derivative $\bar{X}$ in the gradient computation \eqnref{gradient}, we introduce a decorated matrix momentum $M$ to replace $\bar{X}$.
Then the momentum and the gradient are computed by 
\begin{align}
    M_{t+1} & \leftarrow \beta_m M_t + \bar{X}_t, \\
    G_X & \leftarrow M_{t+1} - \frac{1}{2}(X_t X_t^T M_{t+1} + X_t M_{t+1}^T X_t), \\
    M_{t+1} & \leftarrow G_X, \\
    X_{t+1} & \leftarrow R_{\eta}^{G_X}(X_t), \label{Mretraction}
\end{align}
where $M_0 = 0$ and $\beta_m$ is a hyperparameter to be tuned.
Here $R_{\eta}^{G_X}(X_t)$ denotes the retraction operation, which can be substituted by Eq. \ref{qr} - \ref{svd} and \eqnref{Cayley}.
Note that we project the momentum onto the tangent space of Stiefel manifold.

A variant adaptive learning rate is also utilized by 
\begin{equation}
    \eta_{\text{adapt}} = \text{min}(\eta, \alpha_{\eta}/(\left \| A \right \| + \epsilon)),
\end{equation}
where $\alpha_{\eta}$ is a hyperparameter to be tuned and $\left \| A \right \|$ denotes the norm of tensor $A$.
The adaptive learning rate is such that a large learning rate is used when the norm of gradient is small.
These techniques may improve the gradient diffusion problem during the optimization process and accelerate the convergence speed.

\section{Application to MERA}
The multiscale entanglement renormalization ansatz (MERA) is a variational ansatz for the ground state and low-energy excited states of critical quantum spin chains. 
Numerically, it has been used to extract universal information, such as scaling dimensions and operator product coefficients from the critical spin chain Hamiltonian. 
Furthermore, it has found applications in the context of holography \cite{Evenbly2017hyper} and emergent geometry \cite{nozaki2012holographic}. 
\subsection{Review of MERA}
\begin{figure}[htbp]
    \centering
    \includegraphics[width=0.5\textwidth]{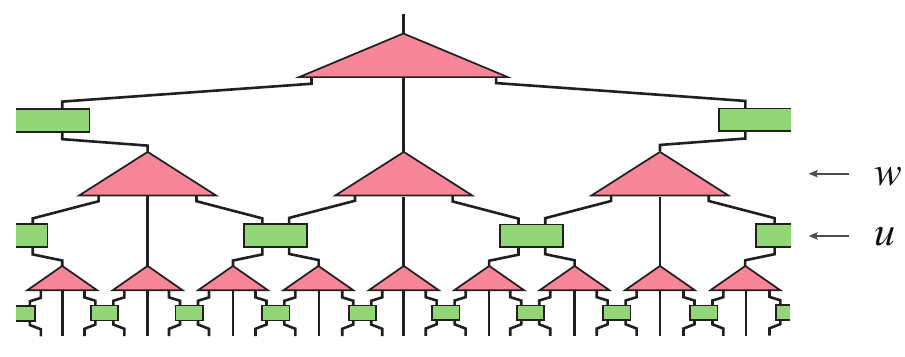}
    \caption{A one-dimensional infinite ternary MERA with isometries $w$ and disentanglers $u$.}
    \label{fig.MERA}
\end{figure}
\begin{figure}[htbp]
    \centering
    \includegraphics[width=0.3\textwidth]{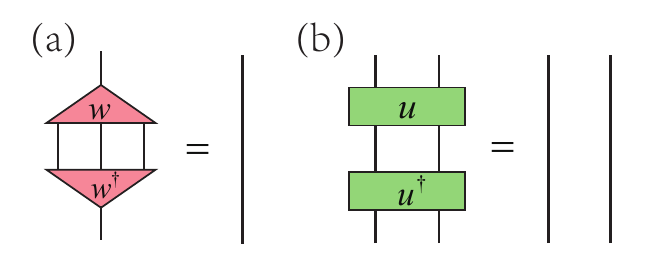}
    \caption{The isometric constraints for isometries $w$ and disentanglers $u$.}
    \label{fig.constraints}
\end{figure}

The MERA $|\psi(u,w)\rangle$ is composed of layers of isometries $w$ and disentanglers $u$ which satisfy the isometric constraints, 
\begin{equation}
    u^{\dagger}u = uu^{\dagger}=I,\quad ww^{\dagger}=I,
\label{constraints}
\end{equation}
which are expressed graphically in \figref{fig.constraints}.
Physically, the MERA can be viewed as a renormalization group flow in the real space. From the bottom upwards, each layer of MERA coarse-grains the quantum state.
The isometric constraints of the tensors are essential from both physical and numerical perspectives. 
From the physical perspective, the isometric constraints keep the norm of the state and retain the causal structure \cite{vidal2007entanglement} under the coarse graining.
From the numerical perspective, the isometric constraints of the tensors ensures that expectation values of the local operators can be computed in polynomial time. 

Depending on the number of bonds of the isometries $w$, there are different types of MERA \cite{evenbly2013quantum} and in this work we focus on the ternary MERA as in \figref{fig.MERA}. We further assume translational invariance, where the $u$ and $w$ tensors are the same in the same layer. We also assume scale invariance, where all $u$'s and $w$'s are the same above a certain number of transitional layers.
Let the number of transitional layers to be $n$, then the variational ansatz is completely specified by $u_{\tau}$ and $w_{\tau}$, where $\tau=1,2,\cdots n+1$ denotes the layers from bottom to top, and $u_{n+1}$ and $w_{n+1}$ constitute the scale-invariant layers.

\begin{figure}[htbp]
    \centering
    \includegraphics[width=0.5\textwidth]{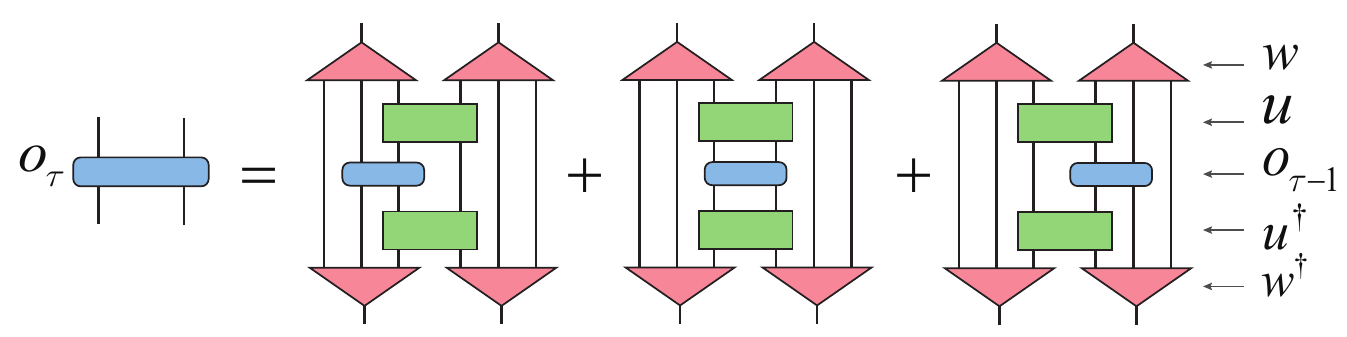}
    \caption{Ascending superoperators ascend the operator $o_{\tau-1}$ at the layer $\tau-1$ to the operator $o_{\tau}$ at the layer $\tau$. They include three different form $\mathcal{A}_L$, $\mathcal{A}_C$ and $\mathcal{A}_R$. The average ascending superoperators $\bar{\mathcal{A}} = (\mathcal{A}_L+\mathcal{A}_C+\mathcal{A}_R)/3$.}
    \label{fig.ascending}
\end{figure}
\begin{figure}[htbp]
    \centering
    \includegraphics[width=0.5\textwidth]{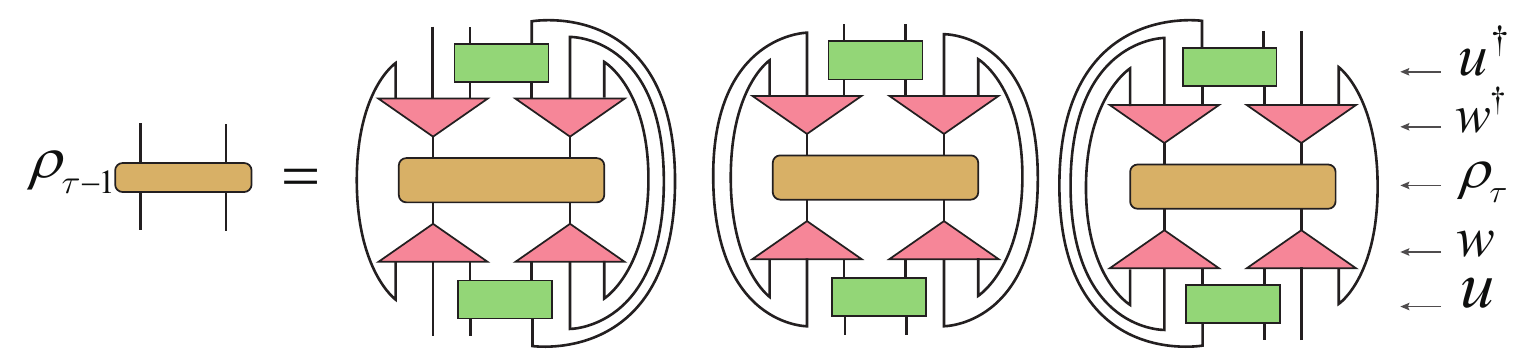}
    \caption{Descending superoperators descend the density tensor $\rho_{\tau}$ at the layer $\tau$ to the density tensor $rho_{\tau-1}$ at the layer $\tau-1$. They include three different form $\mathcal{D}_L$, $\mathcal{D}_C$ and $\mathcal{D}_R$. The average descending superoperators $\bar{\mathcal{D}} = (\mathcal{D}_L+\mathcal{D}_C+\mathcal{D}_R)/3$.}
    \label{fig.descending}
\end{figure}

In order to obtain the ground state, one optimizes the energy function as the loss function,
\begin{equation}
    E = \frac{\langle \psi(u,w)|H|\psi(u,w)\rangle}{\langle \psi(u,w)|\psi(u,w)\rangle}=\langle \psi(u,w)|H|\psi(u,w)\rangle,
\end{equation}
where in the second equality we have used the normalization of the state as a result of the isometric constraints.

It is well known \cite{evenbly2009algorithms} that the computation of the expectation value of local operators in a MERA involves two superoperators, the ''ascending superoperator" and the ''descending superoperator".
The ascending superoperator $\bar{\mathcal{A}}$ transforms local operators $o_{\tau-1}$ at layer $\tau-1$ to local operators $o_{\tau}$ at layer $\tau$, $o_{\tau} = \bar{\mathcal{A}}_{\tau} (o_{\tau-1})$, as shown in \figref{fig.ascending}. The descending superoperator, as the adjoint of the ascending superoperator, transforms the two-site reduced density matrix $\rho_{\tau}$ at layer $\tau$ to the two-site reduced density matrix $\rho_{\tau-1}$ at layer $\tau$, $\rho_{\tau} = \bar{\mathcal{D}}_{\tau} (\rho_{\tau+1})$, as shown in \figref{fig.descending}.

The coarse-graining is such that expectation values of the operator $o_{\tau}$ for each layer $\tau$ are the same
\begin{equation}
    \text{tr}(o \rho) = \text{tr}(o_{\tau} \rho_{\tau}) .
\label{expectvalue}
\end{equation}

At the scale invariant layer $T\equiv n+1$, the two-site reduced density matrix satisfies
\begin{equation}
    \rho_T = \bar{\mathcal{D}}_T (\rho_T),
\label{toprho}
\end{equation}
which is the unique eigenoperator of the average ascending superoperator $\bar{\mathcal{D}}_T$ with eigenvalue $1$. The eigenoperator can be approximately computed by the power method, i.e., repeatedly applying $\bar{\mathcal{D}}_T$ to any initial state until convergent.

To summarize, starting with random isometric tensors $w$, $u$ under the constraints \eqnref{constraints}, the standard computation process of constructing a MERA involves two steps \cite{evenbly2009algorithms}:
\begin{enumerate}
    \item Top-bottom: Computing the density tensor $\rho_T$ at the top layer by solving the eigenoperator of $\bar{\mathcal{D}}_T$, then computing the density tensors $\rho_{\tau}$ for all layers $\tau<T$ from top to bottom by the descending superoperators $\bar{\mathcal{D}}_{\tau}$.
    \item Bottom-Top: From bottom to top, update the isometry and disentangler tensors $w_{\tau}$ and $u_{\tau}$, and compute the Hamiltonian $H_{\tau}$ for all layers $\tau$ by ascending superoperators $\bar{\mathcal{A}}_{\tau}$.
\end{enumerate}

For the updating procedure, we can use Evenbly-Vidal method \cite{evenbly2009algorithms} (see Appendix \ref{app.EVmethod}) or gradient-based methods introduced above.
Repeating the (1) top-bottom and (2) bottom-top process, the tensors $w$ and $u$ in MERA will be constructed.

We note that in differentiable programming only the top-bottom approach is used and the bottom-top approach is computed automatically once the optimization method is determined.

\begin{figure}[htbp]
    \centering
    \includegraphics[width=0.25\textwidth]{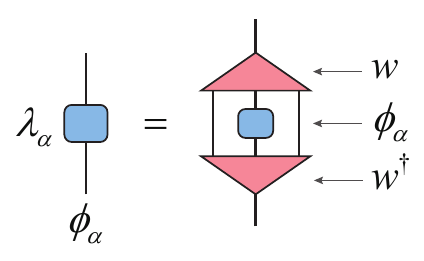}
    \caption{The on-site ascending superoperator acts on the eigenstate $\phi_{\alpha}$ with the eigenvalue $\lambda_{\alpha}$.}
    \label{fig.scaling}
\end{figure}

\textbf{Scaling dimensions}: Once the MERA has been constructed and optimized, we can easily extract the scaling dimensions, which are universal properties of the phase transition that we can obtain easily from MERA.
Take our ternary MERA as example, consider an on-site operator $\phi_{\alpha}$ in scaling invariant layers. 
By the on-site ascending superoperator \figref{fig.scaling}, the operator $\phi_{\alpha}$ is lifted to $\bar{\mathcal{A}}_{\tau}(\phi_{\alpha})$ in the next scale.
The scaling operators are found by the fixed points of the ascending superoperator. 
\begin{equation}
    \bar{\mathcal{A}}_{\tau}(\phi_{\alpha}) = \lambda_{\alpha}\phi_{\alpha}.
\end{equation}
It can be easily shown that the two-point correlator of such scaling operators are
\begin{equation}
    \langle \phi_{\alpha}(3r)\phi_{\alpha}(0) \rangle = \lambda^2_{\alpha} \langle \phi_{\alpha}(r)\phi_{\alpha}(0) \rangle .
\label{correlator}
\end{equation}
The scaling dimensions can therefore be calculated by $\Delta_{\alpha}=-\log_3 \lambda_{\alpha}$. 

\subsection{Auto differentiation}
\begin{figure}[htbp]
    \centering
    \includegraphics[width=0.5\textwidth]{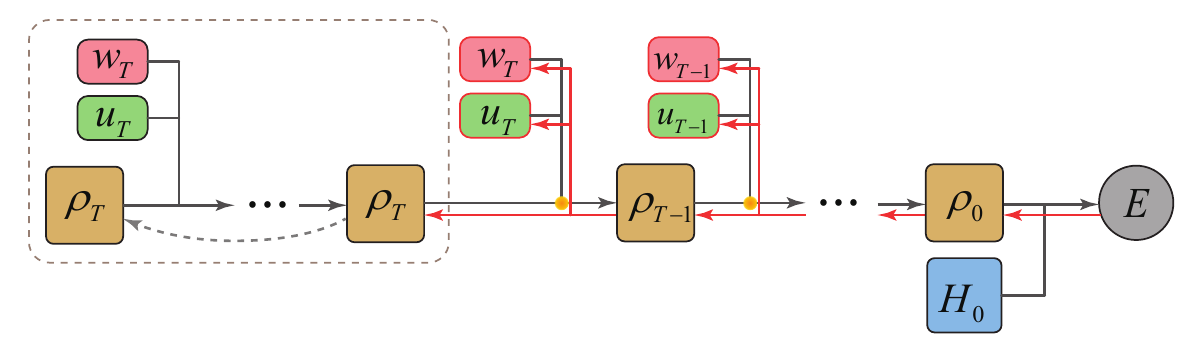}
    \caption{The computation graph for translation invariant MERA. At the left part the density tensor is iterated several times in the scaling invariant layer as the input data. The forward propagation (grey arrow) involves the descending operation with parameters $w$ and $u$ as in \figref{fig.descending}. On the bottom layer, the density matrix is contracted with the Hamiltonian to obtain the energy as the loss function. The backward propagation (red arrow) computes the derivative tensors $\bar{w}_{\tau}$, $\bar{u}_{\tau}$ and $\bar{\rho}_{\tau}$ automatically. Finally the parameters $w$ and $u$ are updated by the gradient optimization method.}
    \label{fig.MERAgraph}
\end{figure}
\textbf{Computation graph}: The computation graph for the loss function of the optimization of MERA is shown in \figref{fig.MERAgraph}.
The forward and backward propagation correspond to the top-bottom and bottom-top processes in the conventional computation, respectively.
The top layer reduced density matrix $\rho_T$ is computed by iteration of the descending superoperator for several layers.
The isometries $w$ and disentanglers $u$ serve as network parameters to be trained.
For each layer $\tau$, the reduced density matrix $\rho_{\tau}$ flows to that of the lower layer $\rho_{\tau-1}$ by the descending superoperators.
Note that the descending superoperators only involve tensor contractions which can be backward propagated automatically.
At the bottom layer, the density matrix $\rho_0$ is contracted with Hamiltonian $H_0$ to obtain the energy $E$ as the loss function $\mathcal{L}$.
Then the derivative tensors $\bar{w}_{\tau}$, $\bar{u}_{\tau}$ and $\bar{\rho}_{\tau}$ of each layer $\tau$ are computed during automatically by backward propagation.
We note that $\bar{w}_{\tau}$ and $\bar{u}_{\tau}$ equal to the environment tensors of $w$ and $u$, respectively.
Finally, due to \eqnref{expectvalue}, we have 
\begin{equation}
    \bar{\rho}_{\tau}=\frac{\partial E}{\partial \rho_{\tau}}=\frac{\partial \text{tr}(H_{\tau} \rho_{\tau})}{\partial \rho_{\tau}}=H_{\tau}.
\label{rhoandH}
\end{equation}

The optimization of $w_{\tau}$ and $u_{\tau}$ can be done by the gradient optimization as in Sec. II. This can be combined with the traditional Evenbly-Vidal algorithm \cite{evenbly2009algorithms} and we compare the performance below.

\textbf{Results}: 
We use the critical one dimensional transverse field Ising model \cite{pfeuty1970one} to test the algorithm.
\begin{equation}
    H_0 = - \sum_r \left( \sigma_x^{[r]} \sigma_x^{[r+1]} + \lambda \sigma_z^{[r]} \right),
\end{equation}
where $\lambda=1$ for the critical point. 

The ground state energy density is known exactly,
\begin{equation}
    E_{\text{exact}} = - \frac{1}{2\pi} \int_{-\pi}^{\pi} \sqrt{(1-\cos k)^2 + \sin^2 k} \,dk = -\frac{4}{\pi} .
\end{equation}

We use the PyTorch framework in Python to realize the auto-differentiable algorithm with the GPU acceleration. 
As comparison of the computation speed, for the optimization of a MERA with $\chi=8$ in $10^4$ iterations using Evenbly-Vidal method, our differentiable programming with GPU acceleration costs $883$ seconds, while the conventional programming without GPU acceleration costs $5109$ seconds. For gradient-based methods the GPU can also accelerate the computation speed. 
Our test platform is a laptop with Intel(R) Core(TM) i7-6700HQ CPU @ 2.60GHz and NVIDIA GeForce GTX 1060 GPU.

\begin{figure}[htbp]
    \centering
    \includegraphics[width=0.48\textwidth]{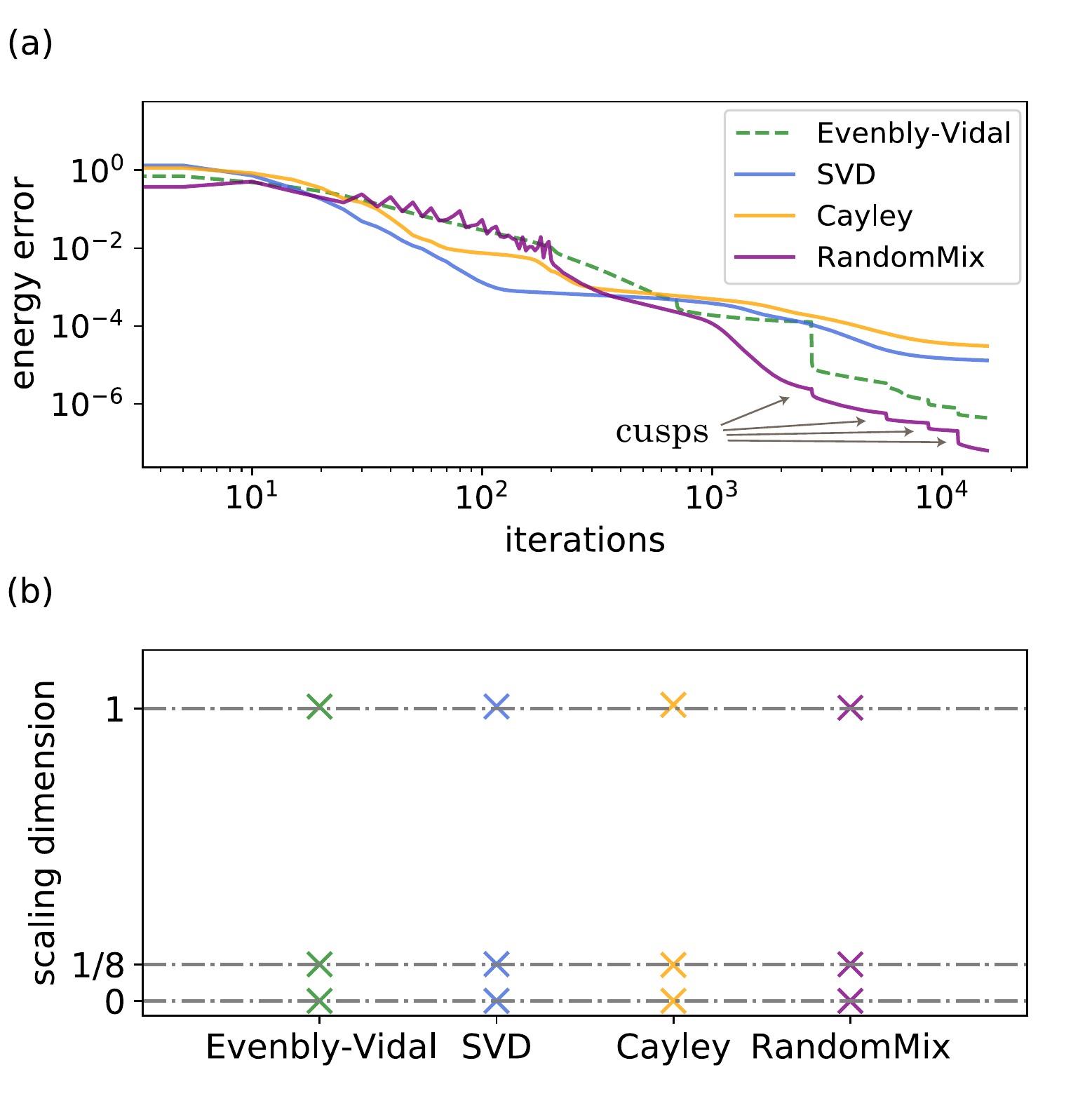}
    \caption{(a) The comparison of optimization curves with Evenbly-Vidal method (green dashed line), SVD method (blue solid line), Cayley method (yellow solid line) and the random mixed method (purple solid line). The max bond dimension $\chi$ is 12 and the number of transitional layers is 3. (b) The computed scaling dimensions with the four method comparing with the exact values (grey dot-dashed line).}
    \label{fig.MERAresults}
\end{figure}

In addition to the Riemannian optimization methods mentioned in \ref{sec:Rieman}, we also propose a random mixed method combining the Evenbly-Vidal method and the gradient-based methods. 
In random mixed method, we use Evenbly-Vidal method as basis method for iterations and replace an iteration by gradient descent methods with SVD or Cayley retractions every 5 iterations.

The error in energy for Evenbly-Vidal, SVD, Cayley and the random mixed methods are shown in \figref{fig.MERAresults} (a).
For SVD and Cayley methods, we introduce dynamical momentum and adaptive learning rate techniques with $\eta=1.0$, $\beta_m=0.9$ and $\alpha_{\eta}=4.0$. 
The learning rate is decayed every 10 iterations with the decay factor $0.999$.
Benefiting from the momentum and adaptive learning rate techniques, the errors of energy can be reduced by more than an order of magnitude. 

It is known that the MERA optimization is vulnerable to be stuck into local minima and gradient diffusion. \cite{hauru2020riemannian} 
In the practice of Ising model optimization, we find that if after the first several hundred iterations the energy error is still larger than a threshold, it will be high probability that the optimization is stuck into local minima. 
In order to reduce the chance of being stuck into local minima and improve the success rate of optimization, we used a resetting mechanism, see Appendix \ref{app.detail} for details of resetting mechanism.
The resetting here means that the network parameters $w$ and $u$ and the two-site reduced density matrix are set to the initial state. 
The number of resetting times can be used as an indicator to show the stability of optimization. High stability means that the optimization has a low possibility to be stuck at a local minima.
We can see from \figref{fig.MERAresults} (a) that for gradient-based methods (SVD, Cayley and the random mixed methods) the energy errors fall quickly at the beginning, meaning that the possibility of being stuck into local minima is lower than the Evenbly-Vidal method. Indeed, we find that these methods hardly trigger the resetting mechanism, while the Evenbly-Vidal method has certain possibility to trigger the resetting mechanism.

\begin{figure}[htbp]
    \centering
    \includegraphics[width=0.5\textwidth]{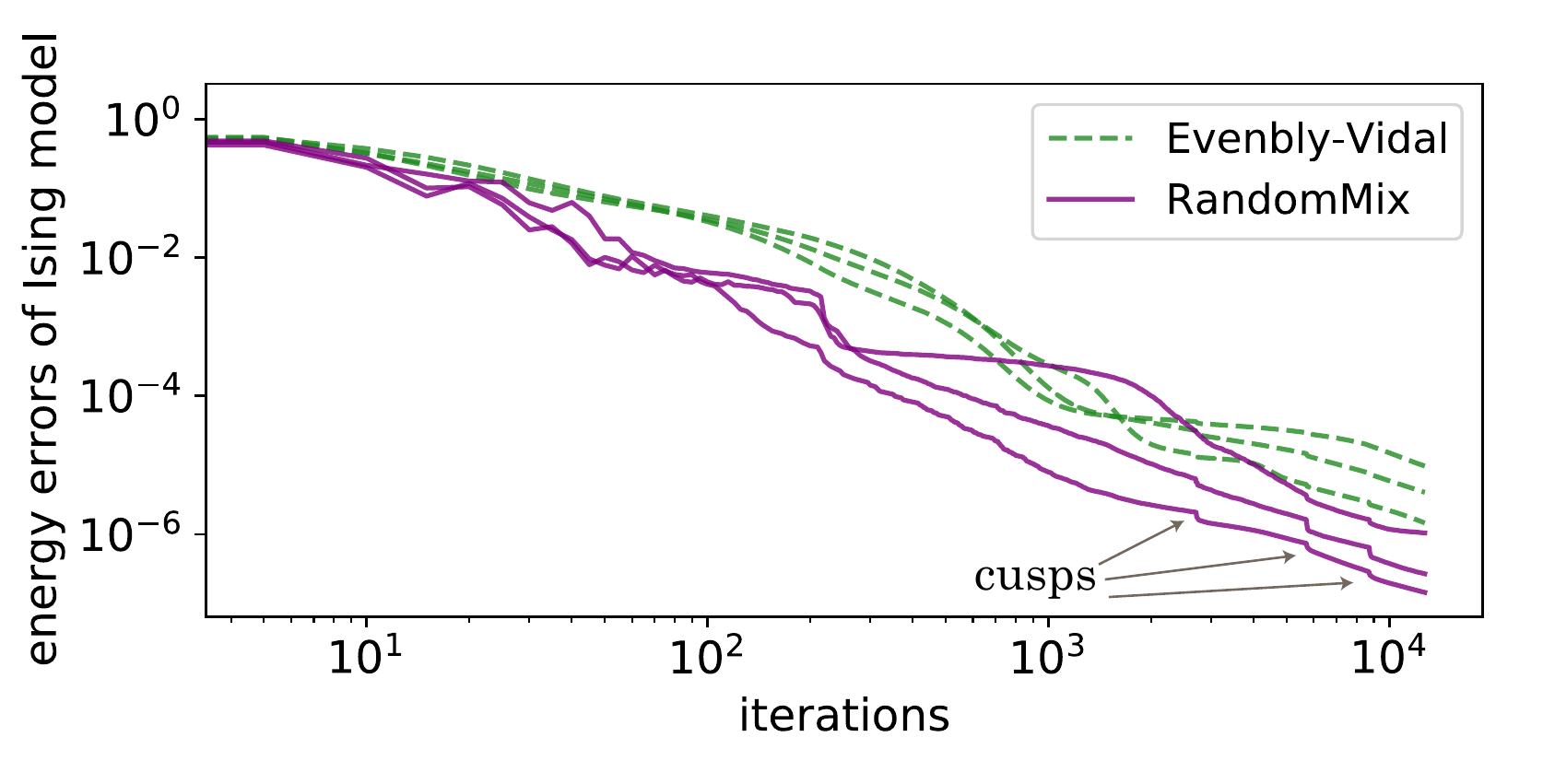}
    \caption{Repeated energy errors with Evenbly-Vidal and random mixed methods with the same setting as \figref{fig.MERAresults}, except for the max bond dimension $\chi=10$ here.}
    \label{fig.MERAlossIsing}
\end{figure}

In order to speed up convergence, we applied a gradually lifting bond dimension trick with the bond dimension $\chi$ increasing from 4 to 12, see Appendix \ref{app.detail} for details of lifting bond dimension trick. 
We can see in \figref{fig.MERAresults} (a) and \figref{fig.MERAlossIsing} that there are some cusps in the errors of energy, corresponding to the bond dimension lifting.

In \figref{fig.MERAresults} (a), we find that although the accuracies of SVD and Cayley methods are not good as Evenbly-Vidal method, the combination of them with Evenbly-Vidal method, corresponding to the random mixed method, has a better accuracy.
We repeat the same optimization process several times for Evenbly-Vidal method and random mixed method as shown in \figref{fig.MERAlossIsing}, finding that generally the random mixed method has better performance in both stability and accuracy.
For other models the performance can also be improved by applying random mixed method, see Appendix \ref{app.othermodel}.

We can provide an explanation for the performance improvement.
It is known that the main difficulty of the optimization comes from the saddle points \cite{pascanu2013difficulty}. 
One idea for escaping saddle points is to introduce fluctuation such as using random input data in Stochastic gradient descent (SGD) \cite{robbins1951stochastic}.
However, in our problem the input data is deterministic (determined by $w$ and $u$ of the top layer in MERA).
Therefore we introduce perturbation of the optimization with Evenbly-Vidal method by randomly choosing SVD and Cayley methods and adding an iteration with this method into the optimization process, as what did in the random mixed method.
Different methods possess different searching way in the parameter space. 
Switching the method meaning the changing of searching way, which can help the optimization process to escape saddle points.
As a result, such a combination of different optimization methods can improve both stability and accuracy benefiting from "shaking up the system". 

The scaling dimensions can also be computed using the optimized MERA. The first three scaling dimensions using the MERA optimized by Evenbly-Vidal, SVD, Cayley and the random-mixed methods are shown in \figref{fig.MERAresults} (b).
As we can see, our optimized MERA could produce good value of scaling dimensions for the first several scaling dimensions. The scaling dimensions of higher order terms usually needs tensor networks with larger bond dimensions. 

\section{application to TNR}
The partition function of a $d+1$-dimensional classical statistical-mechanical models or $d$-dimensional quantum many-body systems can be represented as a $d+1$-dimensional tensor network.
One way of computing the partition function is based on the real-space RG, where the linear size of the tensor network is reduced at each step. 
There are several methods that stood out, including the TRG \cite{levin2007tensor}, HOTRG \cite{hotrg2012}, TNR \cite{evenbly2015tensor,evenbly2017algorithms}, Loop-TNR \cite{loopTNR2017} and Gilt-TNR \cite{giltTNR2018}. Most of the techniques work extremely well in $d=1$, and some of them also work in $d=2$. 
In this work we focus on tensor network renormalization (TNR) in $d=1$, which produces a proper RG flow for a critical system. 
In TNR, there is an optimization over disentanglers and isometries similar to MERA \cite{evenbly2015yield}.
Therefore, the differentiable programming techniques in this work can be applied.
This generalizes previous work where the application of differentiable programming to TRG is discussed \cite{PhysRevX.9.031041}.

\subsection{Review of TNR}

\begin{figure}[htbp]
    \centering
    \includegraphics[width=0.4\textwidth]{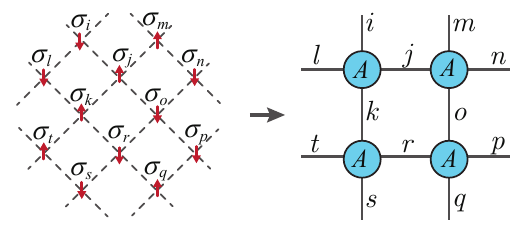}
    \caption{The square lattice of classical spins $\sigma=\pm 1$ rotated by $45^{\circ}$ maps into a tensor network.}
    \label{fig.TNR}
\end{figure}

Here we briefly review the algorithm of TNR. For more details we refer to Ref. \cite{evenbly2017algorithms}.
Consider the 2D classical Ising model on square lattice with inverse temperature $\beta$.
The partition function is 
\begin{equation}
    Z = \sum_{\{\sigma\}} e^{- \beta H({\sigma})},
\label{partition}
\end{equation}
where
\begin{equation}
    H({\sigma}) = - \sum_{\langle i,j \rangle} \sigma_i \sigma_j,
\end{equation}
and $\sigma_i = \pm 1$ is the Ising spin on site $i$.
The partition function is a square tensor network consisting of four-index tensors $A_{ijkl}$ which are located in the center of every second plaquette of Ising spins as in \figref{fig.TNR}, where
\begin{equation}
    A_{ijkl} = e^{\beta (\sigma_i\sigma_j+\sigma_j\sigma_k+\sigma_k\sigma_l+\sigma_l\sigma_i)}.
    \label{Atensor}
\end{equation}
Then the patition function \eqnref{partition} is given by the contraction of the tensor network
\begin{equation}
    Z(\beta) = \sum_{ijk\cdots} A_{ijkl} A_{mnoj} A_{krst} \cdots. 
\label{partition2}
\end{equation}

\begin{figure}[htbp]
    \centering
    \includegraphics[width=0.5\textwidth]{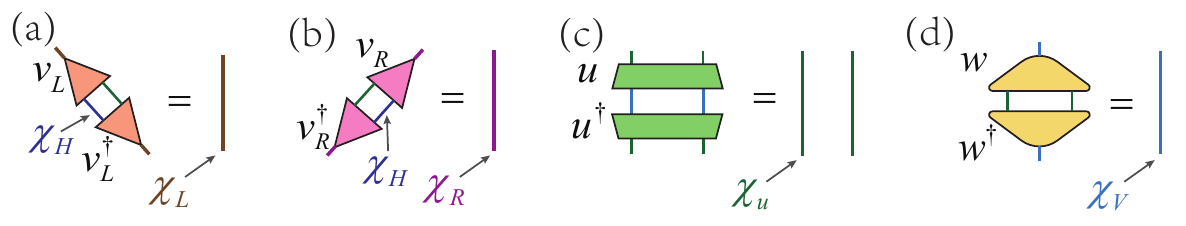}
    \caption{The isometric constraints for (a) the isometries $v_L$, (b) the isometries $v_R$, (c) the disentanglers $u$ and (d) the isometries $w$. The bond dimension $\chi$ of each index are illustrated.}
    \label{fig.TNRconstraints}
\end{figure}

\begin{figure}[htbp]
    \centering
    \includegraphics[width=0.5\textwidth]{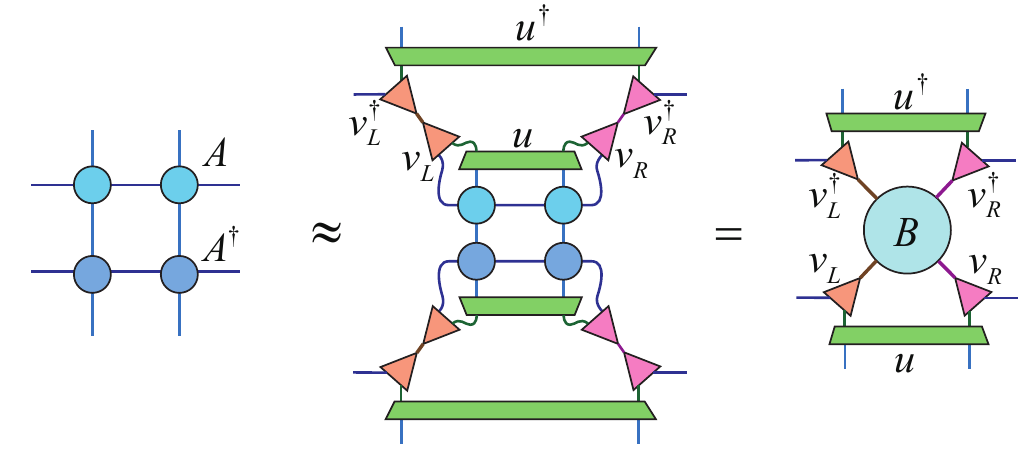}
    \caption{The original sub-network is approximated by a new network by inserting pairs of isometries and disentanglers. The tensor $B$ is defined for convenience.}
    \label{fig.TNRproj}
\end{figure}

\begin{figure}[htbp]
    \centering
    \includegraphics[width=0.3\textwidth]{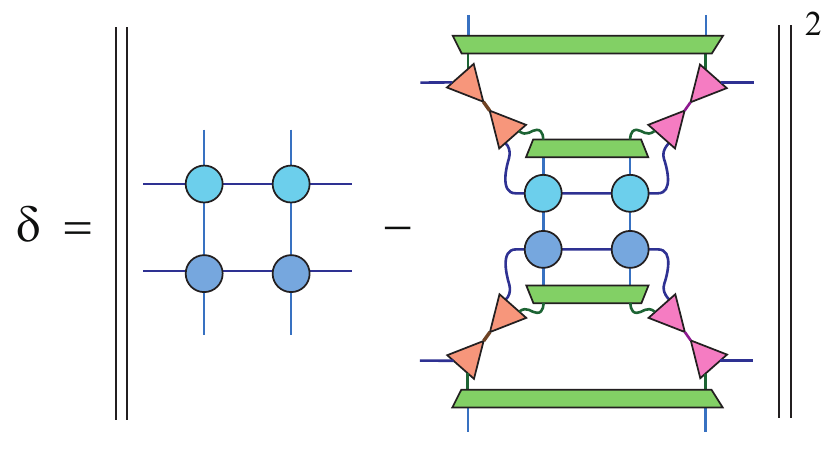}
    \caption{The loss function defined as the approximation error $\delta$.}
    \label{fig.TNRdelta}
\end{figure}

At each step of the coarse-graining, we approximate a $2\times 2$ block of tensors by inserting isometries and unitaries as in \figref{fig.TNRproj}, where the tensors $v_L,v_R,u$ satisfy the isometric constraints as in \figref{fig.TNRconstraints} (a)-(c). 
The tensors $v_L,v_R,u$ are determined by minimizing the approximation error as in \figref{fig.TNRdelta}.

The optimization has been achieved by Evenbly-Vidal method as in Ref. \cite{evenbly2017algorithms}.
Here we use the gradient-based methods like SVD and Cayley methods discussed in this paper.

\begin{figure}[htbp]
    \centering
    \includegraphics[width=0.35\textwidth]{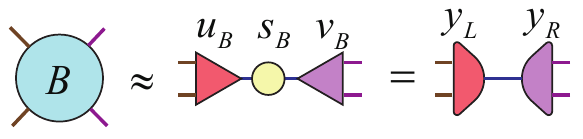}
    \caption{The approximation of tensor $B$ by making singular value decomposition.}
    \label{fig.TNRB}
\end{figure}

\begin{figure}[htbp]
    \centering
    \includegraphics[width=0.45\textwidth]{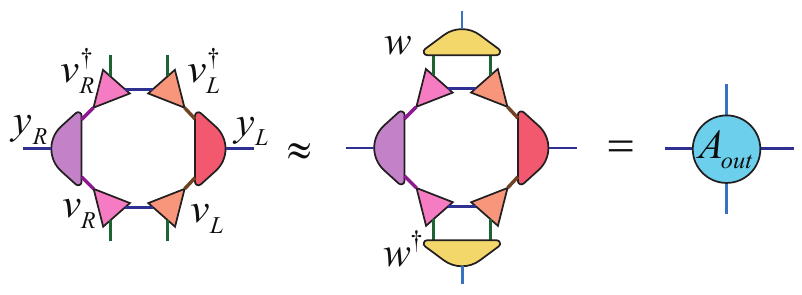}
    \caption{The definition of the coarse grained tensor $A_{out}$.}
    \label{fig.TNRA}
\end{figure}

Then, we contract some of the tensors into a tensor $B$ as defined in \figref{fig.TNRproj}.
Next, a singular value decomposition on the tensor $B$ is performed to obtain the isometric tensors $u_B$, $v_B$ and the diagonal matrix $s_B$ containing singular values, which is truncated to $\chi\times \chi$ by discarding smallest diagonals. 
The diagonal matrix $s_B$ is then splited to make a pair of $\sqrt{s_B}$, and we define new tensors $y_L=u_B\sqrt{s_B}$ and $y_R=\sqrt{s_B}v_B$ as shown in \figref{fig.TNRB}.
The tensor network is now entirely composed of the tensor on the left of \figref{fig.TNRA}.

Finally, we insert the isometry $w$ with the isometric constraint shown in \figref{fig.TNRconstraints} (d) and obtain $A_{out}$, which effectively contain the information of four tensors in the original tensor network. 

We note that the tensors $w$ should be also optimized in order to minimize the approximation errors as in \figref{fig.TNRdelta}.
But here we can apply an alternative method by solving the eigenvectors of the matrix of the first sub-network of \figref{fig.TNRA} with contracting the left and right indices, because of the hermitian property of this sub network. 

Repeating the computation above, the original large tensor network of the partition function can be simplified to one or a few tensors, which can be easily computed.
Also note that at each RG step, the tensor $A$ is divided by the norm of $A$ to prevent the data explosion.

\subsection{Auto differentiation}

\begin{figure*}[htbp]
    \centering
    \includegraphics[width=0.98\textwidth]{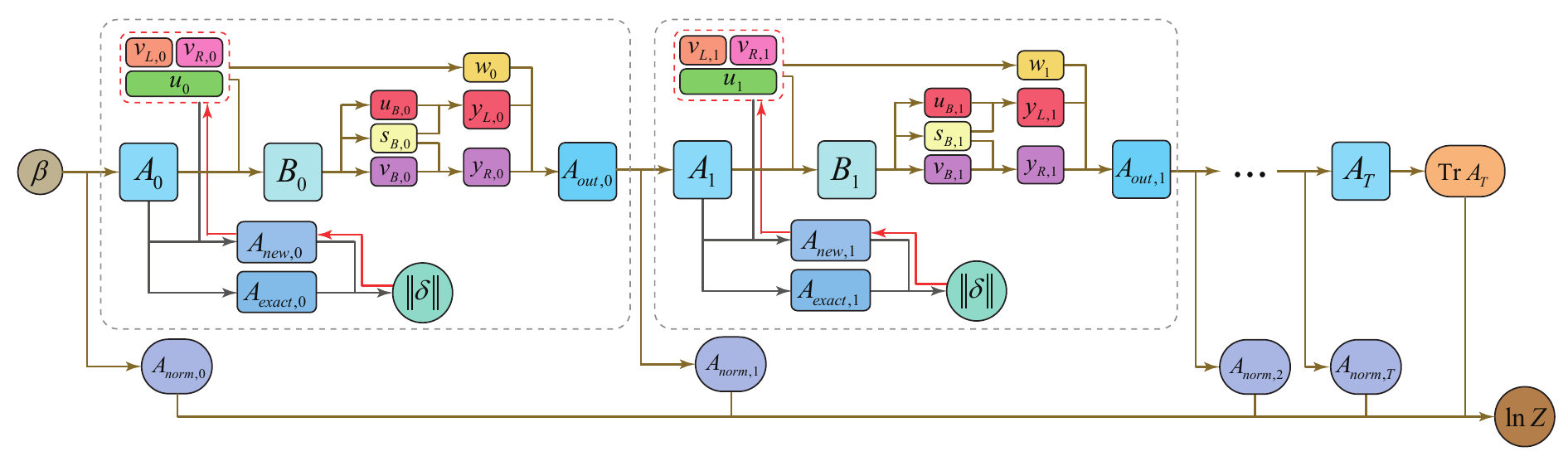}
    \caption{The computation graph for TNR. The grey dotted rounded rectangle indicate one step of TNR computation. The forward and backward propagation processes appear in each coarse-graining step. The gray arrows show the forward propagation path, while the red arrows show the backward propagation path.}
    \label{fig.TNRgraph}
\end{figure*}

\textbf{Computation graph}: Within the differentiable programming framework, we show the computation graph of TNR in \figref{fig.TNRgraph}. 

Given the inverse temperature $\beta$, we first construct the tensor network representation for the system. 
Then we use TNR to coarse grain the tensor network until the network only consists of single or a few tensors.
At each RG step, we use the truncation error as the loss function, and then backward propagate to optimize the parameters $v_L$, $v_R$ and $u$.
We refer to Appendix \ref{app.TNRgraph} for detail description of the computation graph of TNR.

Since the computation graph for the loss function is short, the optimization of parameters $v_L$, $v_R$ and $u$ is easy with the fast convergence.
At the last step where only one tensor $A_T$ remains in the network, we obtain the partition function $\ln{Z}$ by taking the tensor trace of $A_T$ and multiplied by the $A_{norm}$ of all previous steps.

\textbf{Results}: For 2D classical Ising model on an infinite square lattice, the logarithm of partition function is exactly known \cite{onsager1944crystal} 
\onecolumngrid
\begin{equation}
    \centering
    \ln{Z} = \frac{1}{2} \ln{2} + \frac{1}{2\pi} \int_0^{\pi} \ln{ \left( \cosh^2{2\beta} + \frac{1}{\lambda} \sqrt{1 + \lambda^2 - 2 \lambda \cos{2x}} \right)} dx .
\label{lnZ}
\end{equation}
\twocolumngrid

\begin{figure}[htbp]
    \centering
    \includegraphics[width=0.52\textwidth]{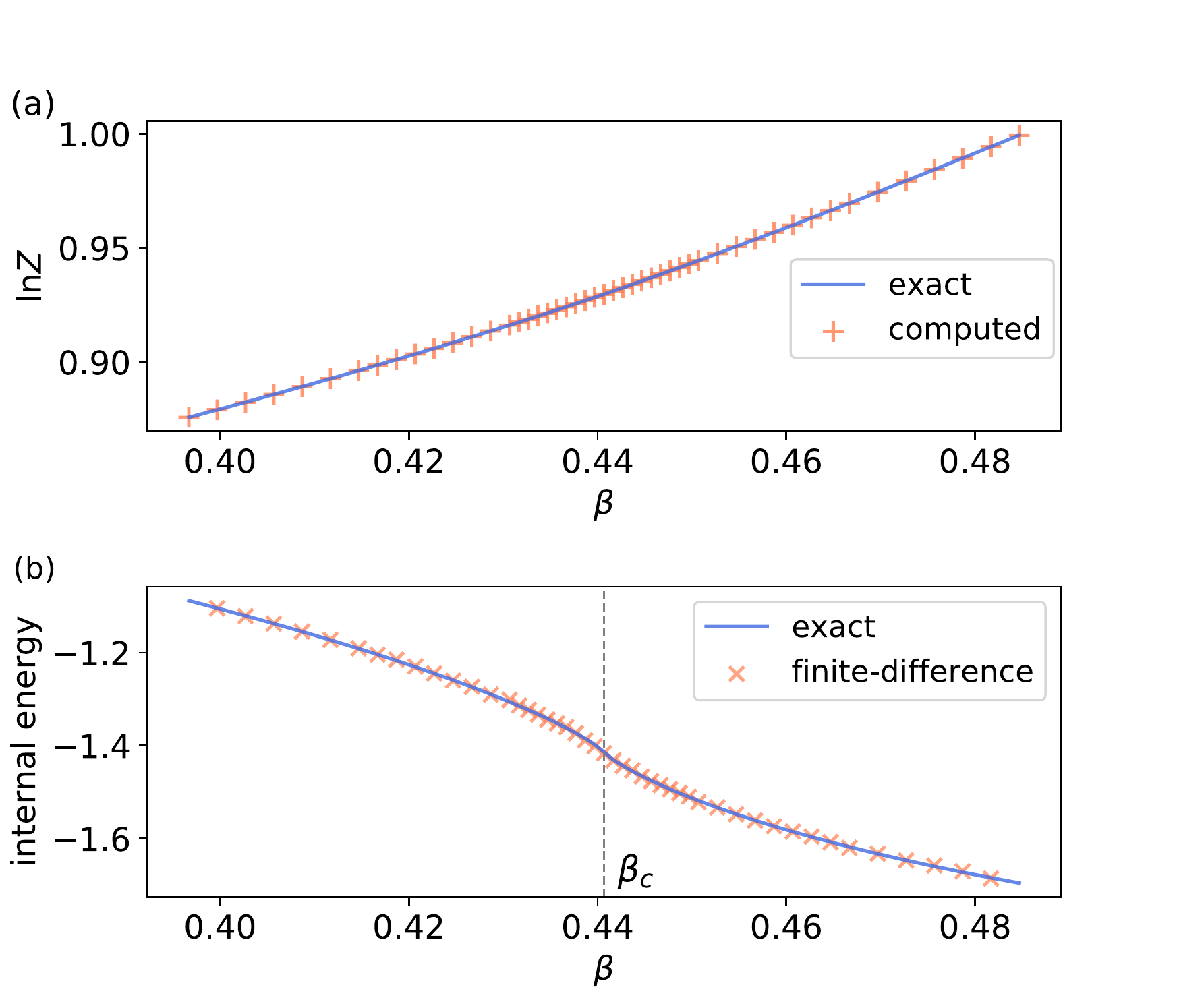}
    \caption{(a) The $\ln{Z}$ of 2D classical Ising model computed by differentiable programming with respected to the inverse temperature $\beta$. The blue line represents the exact value while the orange cross points represent the computed results by differentiable programming. The size of system is $2^8\times2^8$ and the max bond dimension here is $\chi=14$. (b) The internal energy as the function of $\beta$ computed by finite difference method.}
    \label{fig.TNRresults}
\end{figure}

In \figref{fig.TNRresults} (a), we show the computed $\ln{Z}$ accords well with the exact value as a function of the inverse temperature $\beta$ by random mixed method.
We can see that the computed results by differentiable programming fit well with the exact values.

\begin{figure}[htbp]
    \centering
    \includegraphics[width=0.48\textwidth]{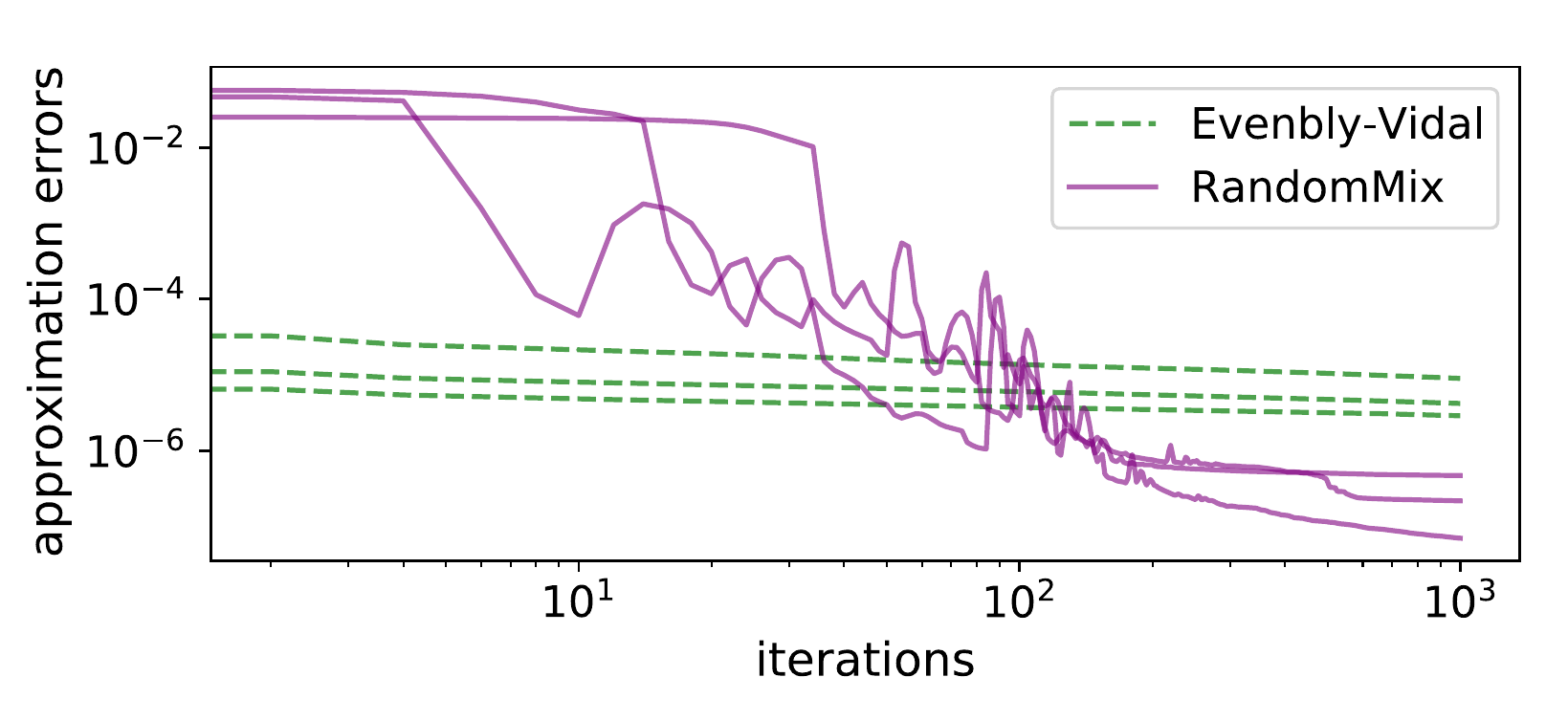}
    \caption{The approximation errors of 2D classical Ising model computed by differentiable programming with Evenbly-Vidal method and random mixed method. Here we take approximation errors of the 3rd, 5th and 7th layers as examples.}
    \label{fig.TNRloss}
\end{figure}

We show the computed internal energy results $E=-\frac{\partial \ln{Z}}{\partial\beta}$ in \figref{fig.TNRresults} (b) by finite difference method. 
We find the random mixed method can also improve the performance of optimization in TNR, as shown in \figref{fig.TNRloss}.

\begin{figure}[htbp]
    \centering
    \includegraphics[width=0.5\textwidth]{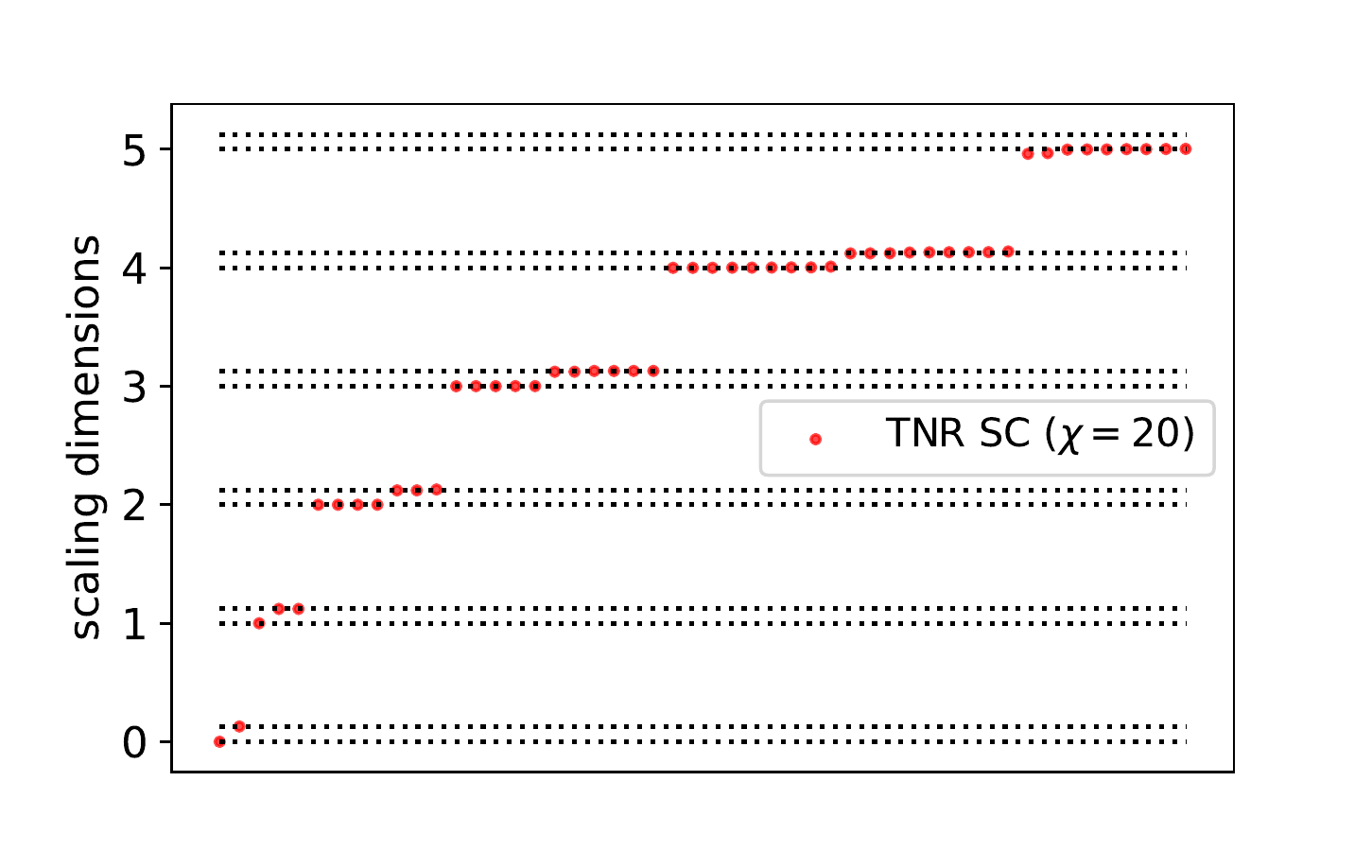}
    \caption{The scaling dimensions of 2D classical Ising model computed by differentiable programming. The dotted gray lines indicate the exact values. The size of system is $2^{10}\times2^{10}$ and the max bond dimension here is $\chi=20$.}
    \label{fig.TNRsc}
\end{figure}

As a comparison of computation speed, for the construction of a $2^8\times2^8$ TNR with $\chi=14$ with $800$ iterations for each layer using random mixed method, our differentiable programming with GPU acceleration costs $708$ seconds, while the conventional programming without GPU acceleration costs 2456 seconds.

\textbf{Scaling dimensions}: Once the TNR has been constructed and optimized, the scaling dimensions can also be extracted easily.
Here we use the transfer matrix technique \cite{hauru2016topological} to compute the scaling dimensions with the critical inverse temperature $\beta_c = \ln{(1+\sqrt{2})}/2$ shown in \figref{fig.TNRsc}.
We can see that with the TNR and transfer matrix technique the computed scaling dimensions have accurate results even at quite higher orders.

\section{Summary}
In this paper, we extend the application of differentiable programming to an important family of tensor networks with isometric constraints such as MERA and TNR.
We show how the MERA and TNR are constructed and optimized within differentiable programming framework by explicitly illustrating the computation graphs of them.
We discuss several gradient-based optimization methods that supplement traditional methods for tensor networks with isometric constraints and show that the performance of optimization can be improved by combining the gradient-based method and traditional methods. 

The differentiable programming in tensor network optimization has several advantages.
(1) Differentiable programming has an unified programming paradigm since we only need to specify the computation graph and the optimization method for the optimization. 
(2) In differentiable programming, the gradient can be computed automatically, which releases the labor on the tedious contraction computation of environment graphs.
(3) Differentiable programming has extensibility and flexibility in practice.
It can be modularized, in which one can easily change and add different network layers and optimization methods.
(4) We can also benefit from the parallel computation with the GPU acceleration.
(5) Last but not the least, differentiable programming can also help with the optimization of arbitrary network networks when no traditional optimization methods are known.

With the merits above, the differentiable programming can be applied to other tensor networks such as higher-dimensional MERA or TNR, which have not yet been implemented previously due to the complexity of the network structures and the high computation expenses.
It is also very interesting and challenging to combine the optimization of tensor data together with the optimization of tensor network's structure \cite{2020arXiv200805437H}. This will serve as a unified method to discover hidden structures of the data. For example, given the measurement results of a quantum state, can machine find the best tensor network structure that represents the data? It has been shown that based on the entanglement data, neural network can establish different hidden geometry (structure) for area law, logarithmic law, and volume law quantum states as an analog of holographic duality \cite{2018PhRvB97d5153Y}. In tensor network studies, it is also known that random tensor networks are closely related to holographic duality \cite{2016JHEP11009H}. Whether different hidden geometry of tensor network can emerge purely based on observation data of different quantum states is both an interesting and fundamental question for physics. Last but not the least, tensor networks with unitary or isometric constraints are closely related to quantum machine learning \cite{cong2019quantum}, as unitary tensors are represented by quantum gates. Our methods can also be applied to simulate and train quantum neural network ansatz for real-world applications \cite{2019arXiv190506352E}. 



\section*{acknowledgements}
We thank Masaki Oshikawa, Yasuhiro Tada, Atsushi Ueda, Jin-Guo Liu, Song Cheng, Yi-Zhuang You and Markus Hauru for helpful discussions. The simulations were performed using the PyTorch \cite{paszke2019pytorch} and we have the open source package "IsoTensor" available on GitHub \cite{geng2021}.
CG was supported by JSPS KAKENHI Grant Numbers JP19H01808. HYH is supported by the UC Hellman
fellowship. YZ is supported by the Stanford Q-Farm Bloch Postdoctoral Fellowship in Quantum Science and Engineering.


\bibliography{MERAref}

%
%

%

\bigskip
\clearpage
\newpage

\appendix
\onecolumngrid
\renewcommand{\appendixname}{Appendix}

\section{Evenbly-Vidal optimization method} \label{app.EVmethod}

\begin{figure} [htbp]
    \centering
    \includegraphics[width=1.0\textwidth]{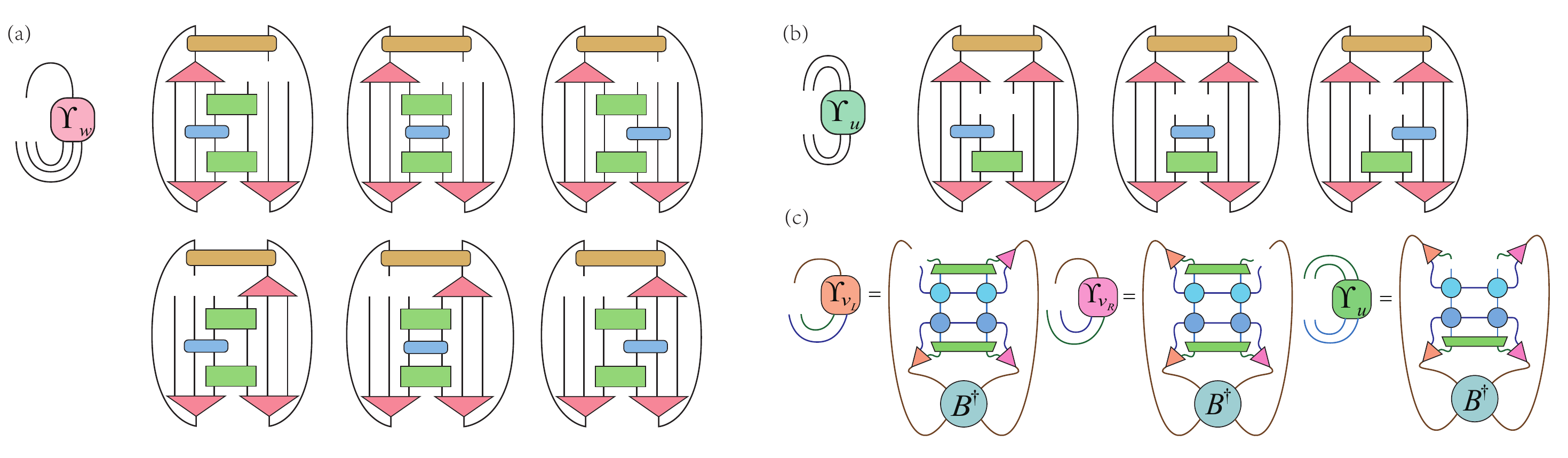}
    \caption{The environment tensors of (a) the isometry $w$ and (b) the disentangler $u$ for MERA, which are the summation of six parts and three parts correspondingly. The environment tensors of $v_L$, $v_R$ and $u$ in TNR are shown in (c).}
    \label{fig.MERAenv}
\end{figure}

In order to optimize the whole tensor network, we need to explain how to optimize a single tensor.
Here we briefly introduce the Evenbly-Vidal optimization method \cite{evenbly2009algorithms,evenbly2017algorithms}.
Taking the isometry tensor $w$ as example, we need to minimize the energy, or to say loss function 
\begin{equation}
    E(w) = \text{tr}(w Y_w).
\label{appendE}
\end{equation}
We temperately regard $w$ and $w^{\dagger}$ as independent tensors and $Y_w$ is the environment of the tensor $w$ shown in \figref{fig.MERAenv} (a).
By applying the singular value decomposition(SVD) on the environment $Y_w = USV^{\dagger}$, the energy $E(w)$ can be minimized if we choose the new isometry tensor
\begin{equation}
    w'=-VU^{\dagger}.
\label{update}
\end{equation}
Then a single step of updating is to replace the old tensor $w$ by the new tensor $w'$.

The remaining quest is to determine the environment of the tensor.
We can automatically obtain the environment tensors by means of differentiable programming discussed in the main text, or drawing and contracting the environment graphs manually.
Here we show the environment graphs for the ternary MERA and TNR used in this paper.

\section{Reducing computation expense of Cayley method} \label{app.Cayleyreduce}
We can reduce the computation of inverting a $n \times n$ matrix to inverting a $2p \times 2p$ ($n>2p$) matrix by Sherman-Morrison-Woodbury formula 
\begin{equation}
    \left(B+\alpha U V^T\right) ^{-1} = B^{-1} - \alpha B^{-1} U \left(I+\alpha V^T B^{-1} U\right) ^{-1} V^T B^{-1}.
\end{equation} 
Define the concatenated matrices $U = [P_X \bar{X},X]$ and $V = [X,-P_X \bar{X}]$ where the square brackets here refer to the matrix concatenation and $P_X = I-\frac{1}{2}X X^T$.
Then we have $A = U V^T$ and the matrix inverse term in \eqnref{Cayley0} becomes 
\begin{align}
    \left(I + \frac{\eta}{2} A\right)^{-1} &= \left(I + \frac{\eta}{2} U V^T\right) ^{-1} \\
    &= I - \frac{\eta}{2} U\left(I+\frac{\eta}{2}V^T U\right) ^{-1} V^T .
\end{align}
And the optimization method of Cayley transform becomes 
\begin{equation}
    X_{t+1} = X_t - \eta U_t \left(I+\frac{\eta}{2}V_t^T U_t\right)^{-1} V_t^T X_t .
\end{equation}

There is an alternative way to accelerate the computation of Cayley transform by the iterative estimation of \eqnref{Cayley0} \cite{wen2013feasible,li2020efficient}.
Note that the fixed point solution of the equation $Y(\eta) = X - \frac{1}{2} \eta A (X + Y(\eta))$ for $Y(\eta)$ is exactly \eqnref{Cayley0}.
By iterating several steps of the equation
\begin{align}
    X_{t+1,0} &= X_t - \eta G_X, \\
    X_{t+1,k+1} &= X_t - \frac{1}{2} \eta A (X_t + X_{t+1,k}) ,
\end{align}
we can update the tensors in the Cayley transform scheme to avoid computing inverse of matrices.

\section{Training details of MERA} \label{app.detail}

\textbf{Resetting mechanism}: In the practice of Ising model optimization of MERA, we find that if after the first several hundred iterations the energy error is still larger than a threshold around $10^{-3}$, it will be highly possible to be stuck into local minima with energy errors $10^{-3} \sim 10^{-4}$. 
In order to reduce the chance of being stuck into local minima and improve the success rate of optimization, we used a resetting mechanism.
To be specific, we first optimize the MERA for $700$ iterations and check whether the energy error is larger than a threshold value $1.5\times10^{-3}$.
If the energy error is larger than the threshold value, we reset the optimization to the beginning, meaning that all the trainable parameters in the network are set to there initial values as well as the associated parameters of optimizers.

We note that the Evenbly-Vidal method in our paper has a little difference from Evenbly and Vidal's original algorithm \cite{evenbly2009algorithms}. In the original algorithm, the update is timely, meaning that once a tensor is updated the next derivative or environment tensor is computed using this updated tensor and so on. within the auto-differentiation framework, it's hard to update tensors timely. That is, we compute all the derivative tensors at once. And then we use these derivative tensors to update all the tensors to be optimized. 

\textbf{Lifting bond dimension trick}: In our MERA optimization we use a gradually lifting bond dimension trick which can speed up convergence and improve the accuracy.
To be specific, we start the computation with a small bond dimension $\chi=4$ in first $200$ iterations. Then we increase the bond dimension to $6, 7, 8, 9, 10, 12$ sequentially by enlarging the tensors' size and padding zero values. 
For the result in \figref{fig.MERAresults}, the lifting bond dimension occurs at the $200, 700, 2700, 5700, 8700, 11700, 15700$ iterations.

\section{methods comparison of MERA for other models} \label{app.othermodel}
In this section we compare the Evenbly-Vidal method and random mixed method for critical Heisenberg XY model and Heisenberg XXZ model \cite{fisher1964magnetism}
\begin{align}
    H_{XY} &= \sum_r \left( \sigma_x^{[r]} \sigma_x^{[r+1]} + \lambda \sigma_y^{[r]} \sigma_y^{[r+1]} \right), \\
    H_{XXZ} &= \sum_r \left( \sigma_x^{[r]} \sigma_x^{[r+1]} + \sigma_y^{[r]} \sigma_y^{[r+1]} + \lambda \sigma_z^{[r]} \sigma_z^{[r+1]} \right)
\end{align}
with $\lambda=1.0$.

\begin{figure} [htbp]
    \centering
    \includegraphics[width=0.45\textwidth]{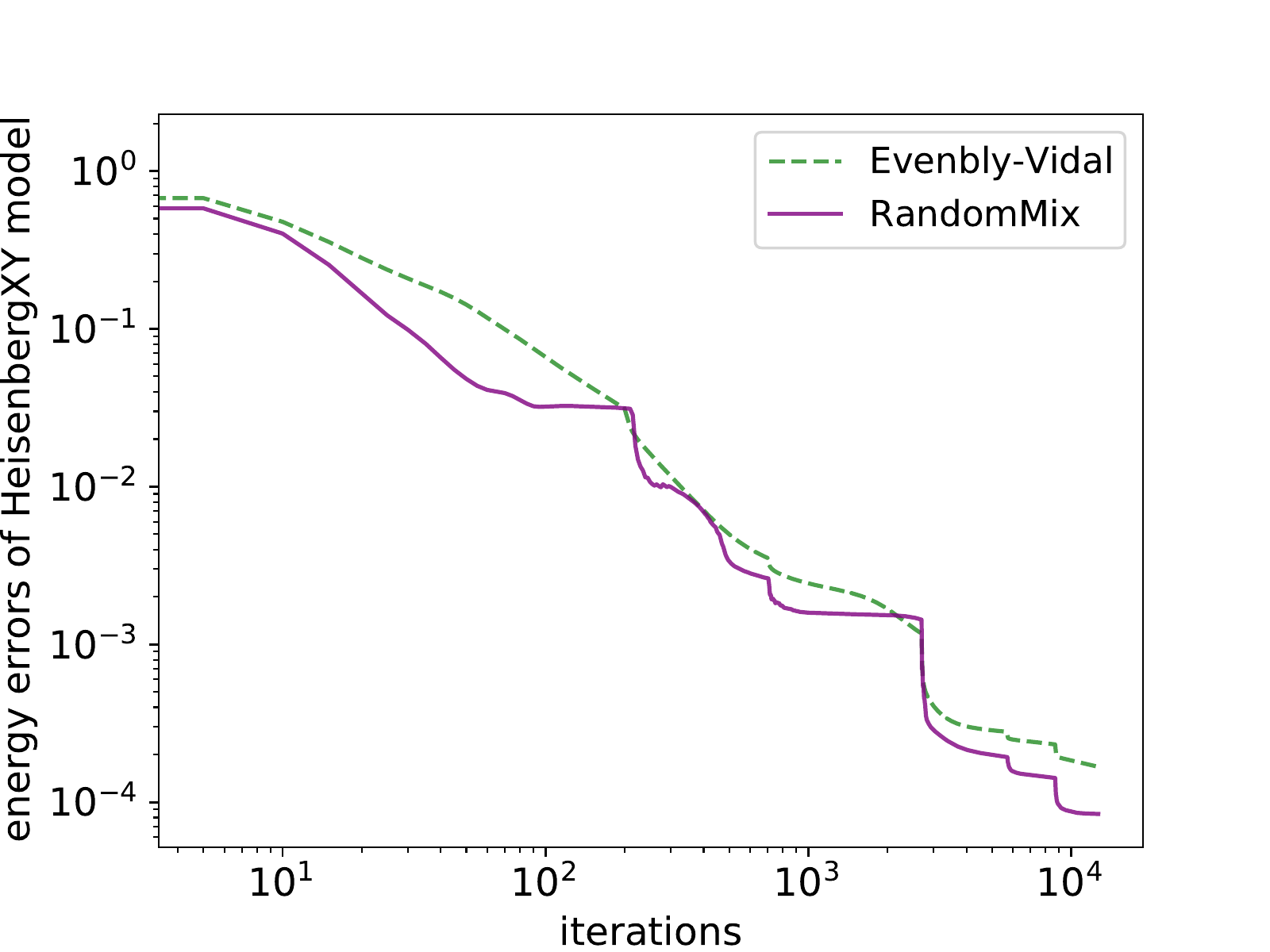}
    \includegraphics[width=0.45\textwidth]{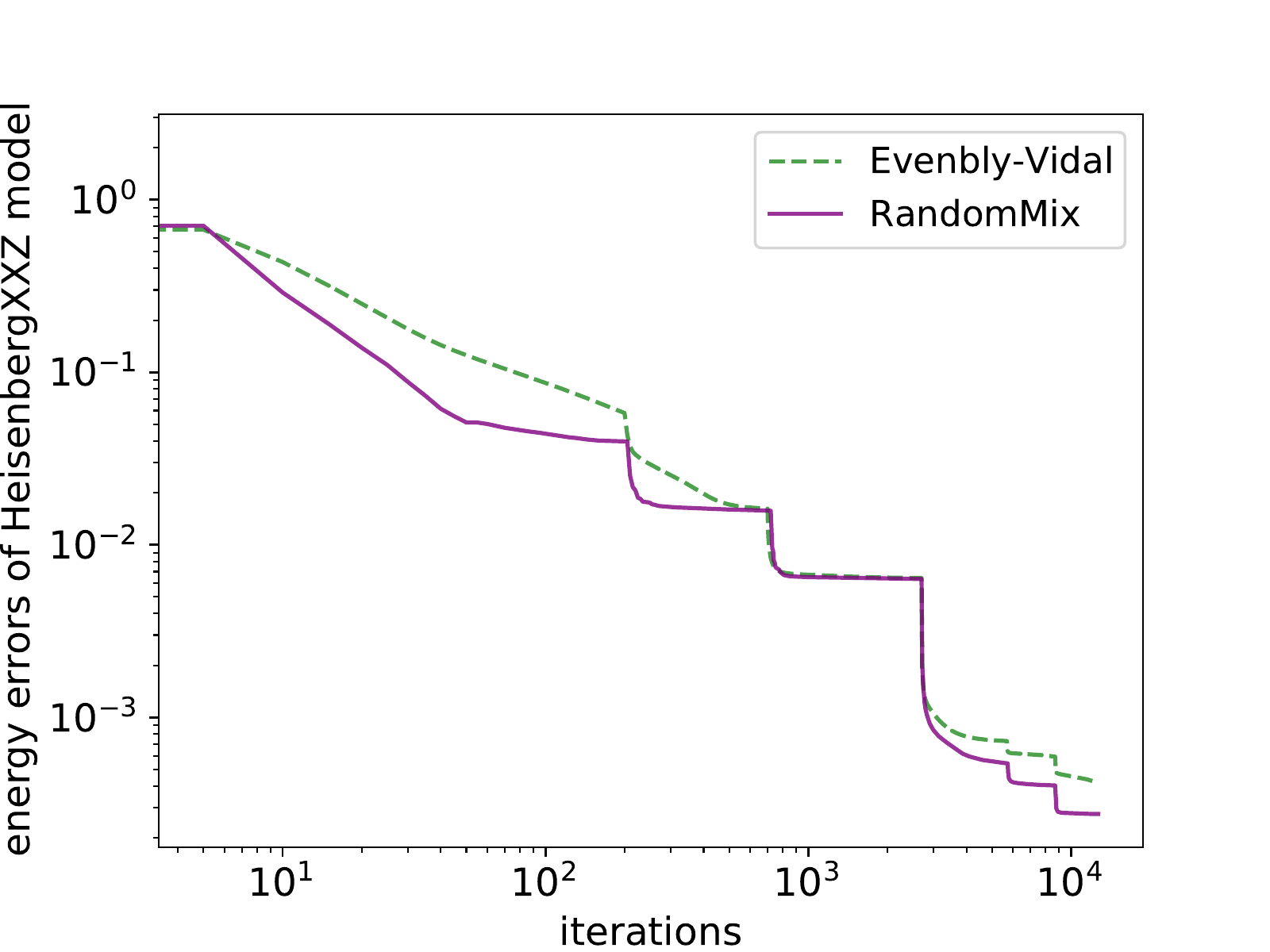}
    \caption{Energy errors with Evenbly-Vidal method and random mixed method for critical Heisenberg XY (left) and XXZ (right) models. The setting of computation is the same as the Ising model in the main text.}
    \label{fig.Heis}
\end{figure}

The results are shown in \figref{fig.Heis}.
We find that for Heisenberg XY and XXZ models the random mixed method can also improve the optimization performance.

\section{Computation graph of TNR} \label{app.TNRgraph}
In this section we explicitly describe the computation graph of TNR used in our paper.
The computation graph for the TNR is illustrated in \figref{fig.TNRgraph} in the main text.

The computation graph starts from the inverse temperature $\beta$ because we would like to investigate the partition function with different $beta$.
Using \eqnref{Atensor} we can represent the partition function as the contraction of $A$s.
In order to prevent the data explosion, the $A$ is renormalized to $A_0$ by dividing a constant number $A_{norm,0}$ which is taken as the $L2$ norm of $A$ here.

The computation graph of TNR for each layer (gray dashed frame in \figref{fig.TNRgraph}) involve three parts.
We take the first layer for example:
\begin{enumerate}[itemsep=0pt,parsep=0pt,label=(\arabic*)]
    \item \textbf{Forward propagation}: Taking a $2 \times 2$ sub-network of $A_0$ we make the approximation by inserting projection operators as in \figref{fig.TNRproj} to obtain the new sub-network $A_{new,0}$.
    Taking a $2 \times 2$ sub-network of $A_0$ we make the approximation by inserting projection operators as in \figref{fig.TNRproj} to obtain the new sub-network $A_{new,0}$.
    Then compare the new sub-network $A_{new,0}$ with the original $2 \times 2$ sub-network of $A_0$ by making contraction of them, we get the approximation error $\left \| \delta \right \|$ as the loss function.
    The forward propagation is shown as gray arrows in \figref{fig.TNRgraph}.
    \item \textbf{Backward propagation}: From the loss function, the derivative tensors of parameters $v_L$, $v_R$ and $u$ are automatically computed.
    With the derivative tensors the parameters can be updated by various methods.
    Iterating the forward and backward propagation the parameters are optimized.
    The backward propagation is shown as red arrows in \figref{fig.TNRgraph}.
    \item \textbf{Tensors renormalization}: Once the parameters have been optimized, we can make the tensors renormalization as described in the main text.
    The tensors renormalization is shown as brown arrows in \figref{fig.TNRgraph}.
\end{enumerate}

By repeating the coarse-graining procedure layer by layer, we finally arrive the top layer and obtain the single tensor $A_T$ in our case. 
Applying the tensor trace on $A_T$ and collecting $A_{norm}$s of every layers, we obtain the partition function $Z$ and its logarithm $\ln{Z}$.

Indeed, the brown arrows in \figref{fig.TNRgraph} also indicate the forward propagation from $\beta$ to $\ln{Z}$. 
We can compute the first-order or even second derivatives of $\ln{Z}$ with respect to $\beta$ corresponding to energy density and specific heat in further works.

\section{Soft-constraint optimization} \label{app.soft}
The soft-constraint optimization is to relax the constraints and allow the tensors $w$ and $u$ to be away from isometric, which is known as the method of Lagrange multipliers \cite{bertsekas2014constrained}.

Define the elementwise average error
\begin{equation}
    T_{err} = \frac{1}{3} \left( \left\langle I-w^{\dagger}w\right\rangle  + \left\langle I-u^{\dagger}u\right\rangle  + \left\langle I-u u^{\dagger}\right\rangle  \right) .
\end{equation}
Here the angle bracket refers to the elementwise average of a matrix.
The value of $T_{err}$ measures how the the tensors $w$ and $u$ away from the isometric constraints.
Adding $T_{err}$ with the energy $E$ we obtain the modified loss function
\begin{equation}
    \mathcal{L}_l = E + \lambda_l T_{err},
\end{equation}
where $\lambda_l$ is the Lagrange multiplier.
Now the constraints are absorbed into the modified loss function and we can optimize the modified loss function $\mathcal{L}_l$ using the usual optimization methods.

The hyperparameter $\lambda_l$ should be tuned manually and the optimization is sensitive to the hyperparameter $\lambda_l$.
But the tuning of hyperparameter $\lambda_l$ is tricky and elusive depending on the systems. 
We introduce a dynamic tuning process of $\lambda_l$ here.
If $\lambda_l$ is too large, the optimization will mainly focus on the constraints and the energy decreasing will become difficult, while if $\lambda_l$ is too small, the optimization will almost ignore the constraints and the energy will keep deceasing.
We use a dynamic tuning strategy:
\begin{enumerate}
    \item Starting with a small $\lambda_l$. After a few iterations switch to a large $\lambda_l$.
    \item When the tensors error $T_{err}$ is lower than a threshold, decrease $\lambda_l$ and the threshold exponentially until to a lower bound.
    \item When the tensors error $T_{err}$ is higher than the threshold, increase $\lambda_l$ exponentially until to a higher bound.
\end{enumerate}

By this tuning strategy, the hyperparameter $\lambda_l$ is tuned automatically depending on the tensors error $T_{err}$ such that the optimization keeps a relative balance between the energy $E$ and the tensor error $T_{err}$.

\begin{figure} [htbp]
    \centering
    \includegraphics[width=0.6\textwidth]{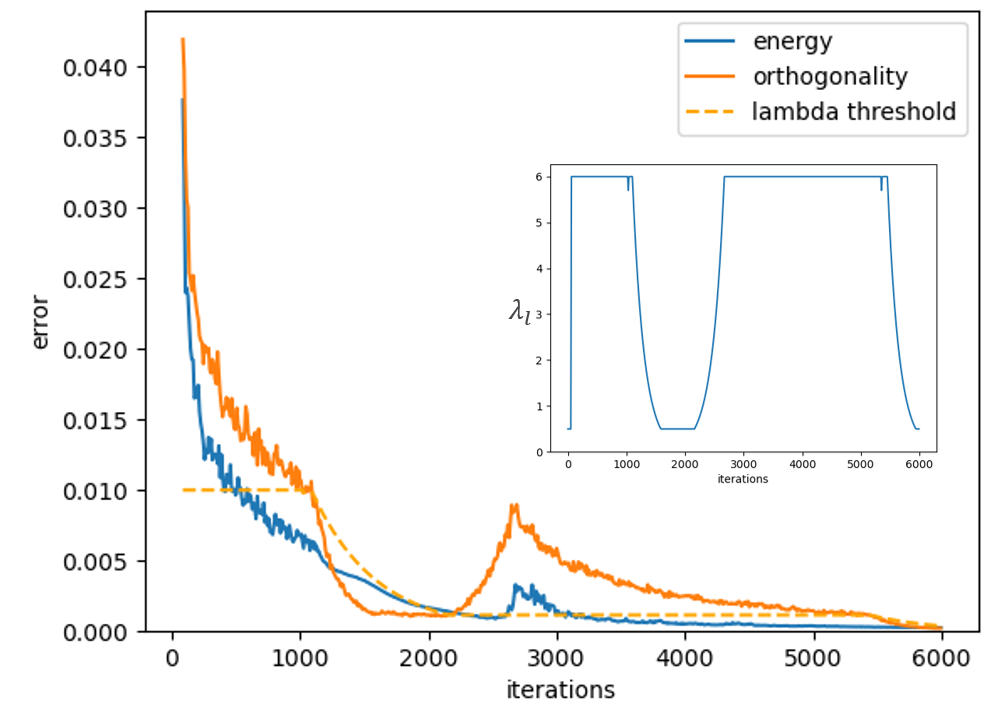}
    \caption{The optimization curves of the soft-constraint method both for the energy error and the orthogonality error. The bond dimension $\chi=8$. The number of transitional layers is 4.}
    \label{fig.soft}
\end{figure}

\figref{fig.soft} shows the optimization curves of the soft-constraint method both for the energy error and the orthogonality error by the Adam optimizer. 
The inner figure shows how the hyperparameter $\lambda_l$ changes during the optimization process.
In this result, the energy error is about $\sim 10^{-5}$, which is larger than what we obtained by Evenbly-Vidal and gradient-based methods in the main text.






\end{document}